\documentclass[pre,showpacs,nofootinbib,twocolumn,showkeys]{revtex4-1}
\usepackage{graphicx}
\usepackage{amsfonts}
\usepackage{amssymb}
\usepackage{amsmath}
\usepackage{array}
\usepackage{dcolumn}
\usepackage{color}
\usepackage{verbatim}
\usepackage{siunitx}

\begin{document}
\title{Simulation studies of the behavior of positrons in a microtrap with long aspect ratio}
\author{Alireza Narimannezhad}
\email{a.narimannezhad@wsu.edu}
\author{Christopher J. Baker}
\author{Marc H. Weber}
\author{Kelvin G. Lynn}
\email{kgl@wsu.edu}
\affiliation{Center for Materials Research, Washington State University, Pullman, WA 99164-2711}

\begin{abstract}
{The charged particles storage capacity of microtraps (micro-Penning-Malmberg traps) with large length to radius aspect ratios and radii of the order of tens of microns was explored. Simulation studies of the motions of charged particles were conducted with particle-in-cell WARP code and the Charged Particle Optics (CPO) program. The new design of the trap consisted of an array of microtraps with substantially lower end electrodes potential than conventional Penning-Malmberg traps, which makes this trap quite portable. It was computationally shown that each microtrap with $50\,\mu m$ radius stored positrons with a density $(1.6\times10^{11}\,cm^{-3})$  even higher than that in conventional Penning-Malmberg traps ($\approx 10^{11}\,cm^{-3}$) while the confinement voltage was only $10\,V$. It was presented in this work how to evaluate and lower the numerical noise by controlling the modeling parameters so the simulated plasma can evolve toward computational equilibrium. The local equilibrium distribution, where longitudinal force balance is satisfied along each magnetic field line, was attained in time scales of the simulation for plasmas initialized with a uniform density and Boltzmann energy distribution. The charge clouds developed the expected radial soft edge density distribution and rigid rotation evolved to some extent. To reach global equilibrium (i.e. rigid rotation)  is to be reached in longer runs. The plasma confinement time and its thermalization were independent of the length. The length-dependency, reported in experiments, is due to the fabrication and field errors. Computationally, more than one hundred million positrons were trapped in one microtrap with $50\,\mu m$ radius and $10\,cm$ length immersed in a $7\,T$ uniform, axial magnetic field, and the density scaled as r$^{-2}$ down to  $3\,\mu m$. Larger densities were trapped with higher barrier potentials.}
\end{abstract}

\maketitle
\tableofcontents

\section{Introduction}
The accumulation and storage of the large quantities of low-energy positrons is becoming increasingly important in different fields. Examples include the study of Bose-Einstein condensation of positronium atoms \cite{Cassidy}, electron-positron plasma in parameter regimes of relevance in astrophysics \cite{GreavesandSurko}, low-energy antihydrogen production and its confinement for long times \cite{Amoretti, Alpha}, studies of the fundamental symmetries of nature \cite{Amoretti}, gravitational interaction of antimatter \cite{Holzscheiter}, and materials science \cite{Schultz}. A more ambitious goal might be the use of antimatter traps to store energy at the maximum possible density per mass unit. Antimatter propulsion of spaceships may well be the only viable method to travel beyond the solar system.

\begin{figure}[!h]
\includegraphics[width=55mm]{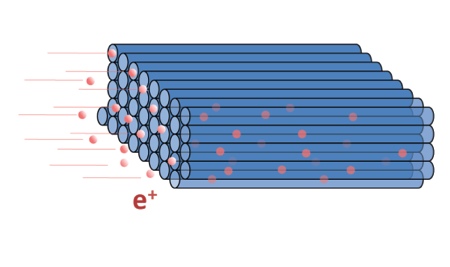}
\caption{\label{array} Schematic configuration of an array of microtraps. The metallic tube electrodes screen the charge in each microtrap. The image is not to scale.}
\end{figure}

Trapping single-component plasmas are the method of choice to accumulate, cool and manipulate a large number of positrons. In principle, these plasmas can be confined by static electric and magnetic fields and be in a state of thermal equilibrium for long periods of time \cite{DavidsonBook}. A number of devices and protocols have been used and proposed to trap antimatter. The Penning-Malmberg (PM) trap \cite{Penning, MalmbergandDriscoll}, because of its ease of construction and versatility, has become the device of choice. To accomplish the goal of energy storage, a fundamental limitation of conventional PM traps must be overcome: the required electrostatic confining potentials rise to large and unpractical values as the charge stored in a PM trap is increased.  A possible solution might be replacing standard traps [aspect ratio $O(10:1)$] with a longer trap to lower the density and avoid the high electrostatic potentials. This is unpractical when the trap length is  $\gg1\,m$. Moreover, increasing the length of the trap can cause some deterioration on the plasma confining time \cite{Eggleston}. The trap can be chopped into segments and aligned in parallel within a single magnet so that the tubes form Faraday cage shields around charge clouds in parallel tubes. The retaining potentials are now fixed as more tubes are added for more positrons. In order to make the overall dimensions feasible (i.e. diameter of the multi-trap $\ll1\,m$) while maintaining high storage densities, each tube diameter should be made much smaller, in the order of tens of microns. MEMS technology as developed for microelectronics may be suitable to fabricate such microtrap arrays. The design of this modified PM trap, of very small space charge potentials in relatively short plasmas compared to the conventional PM traps, reduces plasma heating and weakens the requirements for high uniformity of electrodes and magnetic field and so improves plasma confinement. This design has been proposed by one of the authors (K. G. Lynn) \cite{LynnandGreaves} in order to increase positron storage by orders of magnitude, which consists of an array of microtraps, as shown schematically in Fig. \ref{array}, with large length to radius aspect ratio $O(1000:1)$ and low confinement voltage $O(10\,V)$.  Surko and Greaves \cite{SurkoandGreaves} independently proposed a multi-cell trap, where each cell has a conventional aspect ratio of $10:1$ with a diameter of one centimeter, and its confining voltage is in the order of a few kilovolts. The relative dimensions of length to diameter are not altered very much in contrast to the concept studied here.

Generally, there are two restrictions limiting confinement of large number of positrons in PM traps. One is the Brillouin limit, $n_{B}$  , the maximum density of the plasma confined by a uniform magnetic field  $\vec{B}$, given by \cite{Brillouin}
\begin{equation}\label{Brillouin}
n_{B}=\frac{\epsilon_{0} {\left|\vec{B}\right|}^2}{2m}
\end{equation}
where $m$ is the rest mass of charge particle and $\epsilon_{0}$  is the permittivity of free space. For example, the Brillouin density limits us to  $2.4\times10^{14}\,cm^{-3}$ with use of  $7\,T$ magnetic field.

The second limitation is the space charge potential built up by the confined number of charged particles per length of trap. The space charge potential of the plasma determines the minimum electrical potential required on the end electrodes to confine the plasma in the direction parallel to  $\vec{B}$. In a long, uniform, cylindrical plasma confined with cylindrical metallic electrodes, the space charge potential on the axis of the cylinder is calculated as
\begin{equation}\label{spacecharge}
\varphi_{0}=\frac{qN_{p}}{4\pi  \epsilon_{0} L_{p}}(1+2\ln\frac{R_{w}}{R_{p}}),
\end{equation}
where  $N_{p}$ is the number of particles in the plasma and $q$  is the charge of each particle. For example, with $N_{p}=1.24\times10^{13}$ , a length, $L_{p}$, of $10\,cm$ and a radius,  $R_{p}$, equal to  $R_{w}/\sqrt{3}=1\,cm$ ($R_w$ is the radius of the microtrap), the plasma  has a positron density of $3.95\times10^{11}\,cm^{-3}$.  The required minimum end electrode potential is then $364\,kV$ based on Eq. (\ref{spacecharge}). The array of microtraps, which is showed earlier, circumvents these large space charge potentials because the metallic electrodes screen the charge in each microtrap. Extending the length of the trap by a factor of $10000$ lowers the potential to $36.4\,V$. $10000$ conducting wall tubes of $10\,cm$ length each in parallel also require only $36.4\,V$. Shrinking the radii of the tubes, while maintaining the ratio of $R_w/R_p$ does not cost extra potential.

\begin{figure}[!h]
\includegraphics[width=75mm]{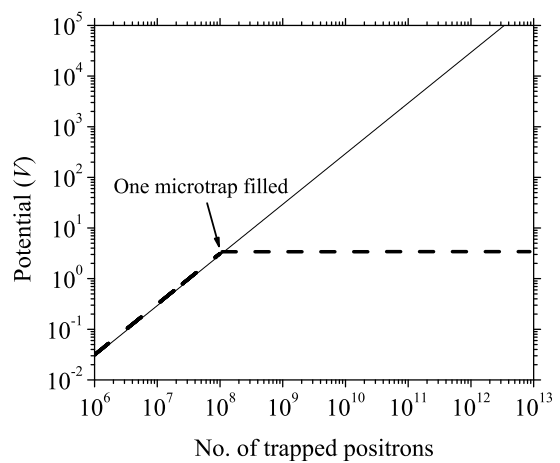}
\caption{\label{seqfill} Space charge potential developed in a microtrap array, the dashed curve, compared to a conventional PM trap, the solid curve, assuming that the array is filled sequentially up to $10^8$ positrons per tube, calculated from Eq. (\ref{spacecharge}) when  $R_p=R_{w}/\sqrt{3}$, $R_w=50\,\mu m$ and $L_p=10\,cm$.}
\end{figure}

The space charge potential of a microtrap array is compared to a conventional PM trap in Fig. \ref{seqfill}. The maximum space charge on axis for each microtrap is chosen as $3.75\,V$. Note that the magnitude of confining electric barrier is linearly increasing with the number of trapped positrons in conventional PM trap. The idea to design the trap which consists of an array of microtraps is mainly to avoid the financial cost of an exceedingly high repulsive electric barrier and improve the portability of these traps.

In order to study the storage capacity of positrons in micortraps, one should trace the plasma behavior toward the equilibrium. A non-neutral plasma in a PM trap with rotational symmetry along the $z$ axis is to reach an equilibrium configuration, in which there is no torque on the plasma and no transport across the magnetic field and it will be confined for an infinite time in principle. If we neglect the charges radiation, the cylindrical symmetry of the trap potential implies that the total angular momentum is conserved,
\begin{equation}\label{angularmomentum}
\sum\limits_{i=1}^{N} p_{{\theta}_{i}}=const.
\end{equation}

Of course, as the cylindrical symmetry is broken in the trap due to the fields and construction errors, the particles are lost in time. However, the time scale for that is normally long compared to the time required for the charges to attain the thermal equilibrium \cite{Dubin}. In an equilibrium configuration, which can be obtained up to the Brillouin density, the inward Lorentz force is balanced with the outward centrifugal force, pressure, and electric force on the plasma. The plasma rotates rigidly along the direction of the magnetic field. If one considers the Hamiltonian in the rotating frame of plasma, the effective trap potential is calculated as  \cite{Dubin}
\begin{equation}\label{effectivepotential}
q\phi_R=q\phi_T+m\omega(\Omega_c-\omega)r^2/2,
\end{equation}
where $\omega$ is the plasma rotating frequency, $\Omega_c$ is the cyclotron frequency, and $\phi_T$ is the trap electric potential in the absence of the plasma. So we have
\begin{equation}\label{effectivepotential2}
q\phi_R=q\phi_T-m\omega^2r^2/2+q\omega r^2 B/2,
\end{equation}
in which the last term is the potential due to the plasma rotation through the magnetic field. Note that the magnitude of $q\phi_R$ increases form center toward the end electrodes since $q\phi_T$ is increasing in this direction. To ensure that all the plasma is confined, the effective trap potential should be also increasing in $r$ by choosing high enough magnetic field although the first term, $q\phi_T$, decreases in outward direction. The effective trap potential acts as a potential well to confine the particles. Particles should use their energies to climb up this potential well. In other words, it compels the density to be exponentially small at the end electrodes and at large radii. Confinement requires the walls to be located beyond the radius where the density drops to zero.

Any externally imposed electrostatic field is Debye shielded out at the state of equilibrium. Hence, the density is almost constant at the inside region \cite{Davidsonbook1990},
\begin{equation}\label{Debyeshield}
\phi_p+\phi_R\approx const.
\end{equation}
where $\phi_p$ is the space charge potential. Taking the Laplacian and using the Poisson's equation and Eq. (\ref{effectivepotential})  we can write
\begin{equation}\label{rotationfrequency}
nq^2/\epsilon_0 \approx 2m\omega(\Omega_c-\omega),
\end{equation}
by which the rotation frequency of the plasma, $\omega$, is obtained at the equilibrium. The same value is calculated for $\omega$ by writing the force balance in radial direction at the inside region where the density is almost constant out (constant pressure),
\begin{equation}\label{forcebalanceinside}
nq\omega r B-nm{\omega}^2 r\approx nqE_r,
\end{equation}
where $E_r=nrq/2\epsilon$ is radial electric field. $\omega$ is obtained as
\begin{equation}\label{rotationfrequency2}
\omega \approx \frac{\Omega_c \pm \sqrt{\Omega_c^2-2\omega_p^2}}{2},
\end{equation}
where $\omega_p=\sqrt{nq^2/{\epsilon_0 m}}$ is the plasma frequency. The two possible solutions for $\omega$ are real numbers only for densities below the Brillouin limit, $\epsilon B^2/2m$.

\section{Simulation model}
\begin{figure}[!h]
\includegraphics[width=60mm]{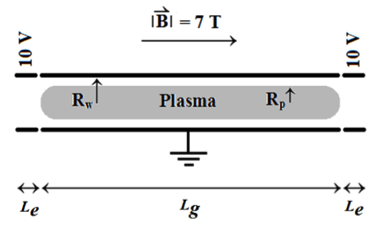}
\caption{\label{model} The schematic geometry of a microtrap and a plasma. The image is not to scale.}
\end{figure}

The modeling geometry and simulation parameters are explained in this section. A schematic of one microtrap modeled in our simulations is shown in Fig. \ref{model}. It is composed of a central perfectly electrically conducting grounded tube and two end electrodes. The trap is cylindrically symmetrical and the potentials on the end electrodes are constant. The gap between the tube and the end electrodes is comparable to the mesh size so the electrodes with different potentials are as close as possible without touching. The tube is immersed in a uniform, constant, axial magnetic field. Typical modeling parameters are listed in Table I.

\begin{table}[!h]
\centering
\caption{\label{table:1}The modeling parameters of the simulation. Those without magnitudes are varied.}
\begin{tabular}{lc}
\toprule
Modeling parameters        &     Symbol and/or magnitude \\
\hline
Magnetic field              & $B=7\,T$ \\
Grounded central tube length & $L_g= 2$ to $360\,mm$ \\
End electrode length & $L_{e}=1$ to $10\,mm$ \\
The radius of the microtrap & $R_w=3$ to $50\,\mu m$ \\
Main tube potential & $V_w=0\, V$ \\
End electrodes potential & $V_e=10$ to $500\,V$ \\
Initial plasma radius & $R_p=R_{w}/\sqrt{3}$ \\
Initial plasma length & $L_p=0.27$ to $9.998\,cm$ \\
Initial space charge on $z$ axis & $\varphi_0=0.007$ to $0.375\,V$ \\
The initial plasma density & $n_0=2.1\times10^{10}$ to $9.76\times10^{13}$ \\
Initial plasma temperature & $T_0=0.025$ to $0.5\,eV$ \\
\toprule
\end{tabular}
\end{table}

\begin{figure}[!h]
\includegraphics[width=85mm]{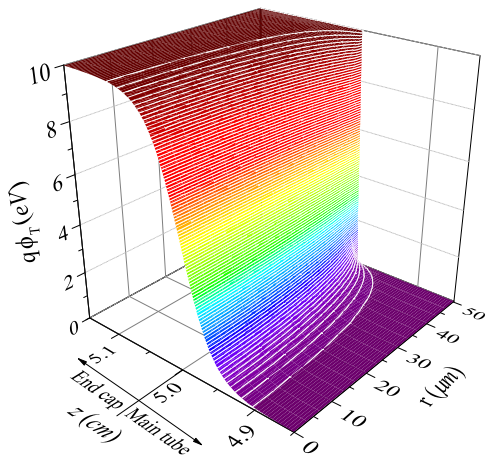}
\caption{\label{PhiEmpty} The potential energy, $q\phi_T$, inside a microtrap of $50\,\mu m$ radius in the absence of the plasma, shown on the region where the main tube meets the end electrode.}
\end{figure}

\begin{figure}[!h]
\includegraphics[width=60mm]{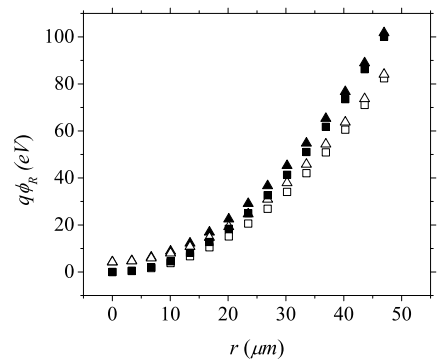}
\caption{\label{effectiveV} The effective trap potential as a function of $r$ for two possible plasma rotation frequencies when density is $10^{12}\,cm^{-3}$, at $z=0$, $\square$ and $\blacksquare$, and at $z=L_g/2=5\, cm$, $\vartriangle$ and $\blacktriangle$.}
\end{figure}

The axial confinement of a symmetric plasma is most worrisome at its longitudinal axis where the space charge potential is highest. The bias voltage is simply chosen by calculating the space charge and taking the energies of the particles into account. As for the radial confinement, one can calculate the effective trap potential to apply a sufficiently high magnetic field in order to trap a certain density. Here, the magnetic field strength is fixed at $7\,T$ in all simulations. We investigate how much density can be trapped in the microtrap with a certain radius and end electrodes potential. Calculating $\omega$ from Eq. (\ref{rotationfrequency}) and applying into Eq. (\ref{effectivepotential2}), we can figure out the effective trap potential in the state of equilibrium and adjust the depth of potential well in radial direction. Fig. \ref{PhiEmpty} shows the potential energy, $q\phi_T$, due to the voltages maintained on the electrodes of a microtrap of $50\,\mu m$ radius in the absence of the plasma. The potential is zero and constant in $r$ far from the end electrodes and becomes decreasing in $r$ as nearing them.  Two different cross sections in Fig. \ref{PhiEmpty} are considered for calculating the effective trap potential in Fig. \ref{effectiveV}, one at $z=0$, the center of the trap, and the other one at $z=L_g/2=5\, cm$, where the grounded tube meets the end electrode. For a density of $10^{12}\,cm^{-3}$ as an example, the effective trap potential is illustrated in Fig. \ref{effectiveV} for two possible rotation frequencies.

\begin{figure}[!h]
\includegraphics[width=75mm]{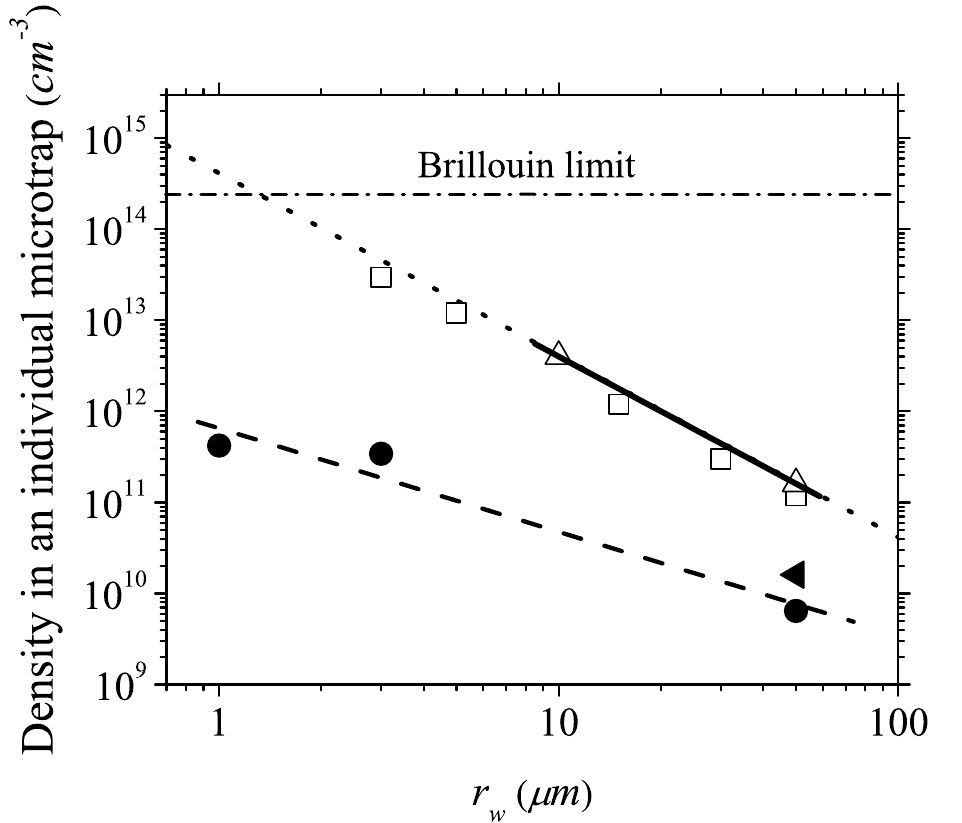}
\caption{\label{analytical}The density of the plasma as a function of the radius of the microtrap when  ${R_w}/{R_p}=\sqrt{3}$. The results from short cases S1-S5 of WARP are shown, $\square$. The cases W4 and W11 of WARP, $\vartriangle$, and also the cases S6-S8 of CPO, $\bullet$, are fitted with $\ln n=a+b\ln R_w$  , the solid line and the dashed line respectively, enabling the comparison with the analytical expectation, the dot line. The WARP data calculates  $a=14.6$ and $b=-2$ , and the CPO data calculates $a=11.812\pm 0.273$  and  $b=-1.136\pm 0.268$. Well studied case of CPO (the case C2),$\blacktriangleleft$, for $50\,\mu m$ radius microtrap is discussed in section V.}
\end{figure}

Modeling simulations were carried out with two different computational tools: WARP, a code used extensively in plasma physics \cite{WARP}, and Charged Particle Optics program (CPO) \cite{cpo}. Simulations conducted with $50\, \mu m$ and  $10\,\mu m$ radius microtraps are discussed in detail in Section III. All the parameters beside those shown in Table I are listed in Tables II and III. Shorter time modelings, cases S1-S8, presented in Fig. \ref{analytical} were done using given parameters. The trapped particles are then extrapolated in time to get the values of lost particles at infinite time. In order to get the WARP simulations done in timely manner, modeling required to shorten the microtrap length as its radius decreased. It is discussed in section III whether the length of the trap affects the results. All of the WARP short simulations used parallel processing of eight  $2.53\,GHz$ Intel\textregistered Xeon\textregistered CPUs, while for long runs, cases W1 to W11, modeling used 9 nodes of WSUÕs high performance computer \cite{HPC}; totaling 108 Intel\textregistered Xeon\textregistered CPUs each running at  $2.4\,GHz$. All the CPO simulations used a  $3.2\,GHz$ Intel\textregistered Xeon\textregistered CPU. The startup parameters of the plasma played a vital role in reducing the computational effort. The closer the initial guesses of the plasma density distribution and spatial distribution, the faster the codes probed the long term evolution of the plasma in a given trap geometry. The major simulation parameters such as time step and mesh size were chosen carefully. Discordant values of them cause a large numerical instability as the Courant-Friedrichs-Lewy (CFL) condition \cite{CFL} requires that the particles must not move further than one mesh size during one time step in the plasma simulation.

The $50\, \mu m$ and  $10\,\mu m$ radius microtraps are the focus of the paper. The plasmas are still large enough that the free energy is dominated by the bulk plasma. Nonetheless, as the trap size shrinks much more, the number of particles gets so small that the trapped could not be called plasma anymore. The term plasma is referred to a weakly correlated cloud of charges when it is large compared to the Debye length, $\lambda_D=(\epsilon_0 kT/ nq^2)^{1/2}$. A weakly correlated cloud has a coupling parameter, $\Gamma= e^2 / a k T$, much smaller than 1. Here, a is the Wigner-Seitz radius where $4 \pi n a^3 /3=1$ \cite{Dubin}. 

\begin{widetext}

\begin{table}[!h]
\centering
\caption{\label{table:2}The parameters used in WARP simulation.}
\begin{tabular}{c*{14}{c}r}
\toprule
&\multicolumn{10}{c}{Modeling parameters} \\
\cline{2-11}
Case & $R_w $ & $\Delta t \footnote[1]{Time step} $ & $\Delta R\footnote[2]{Mesh size}  $ & $PW\footnote[3]{Positron weight: the number of real particles that each simulation macro-particle represents.} $ & $T_0 $ & $\;\;\;\varphi_0\;\;\; $ & $V_e$ & $n $ & $L_e $ & $L_g $ & $L_p$ & $\lambda_{D}/R_w$ & Total time & Run time  & $\;\;e^{+}$ lost  \\
No. & $(\mu m)$ & $(ps)$ & $(\mu m)$ & & $(eV)$ & $(V)$ &$(V)$ & $(cm^{-3})$ & $(mm)$ & $(cm)$ & $(cm)$ && $(\mu s)$  & $(hr)$ & $(\%)$ \\
\hline
S1	& 3&	0.5&	0.24&	5&	0.025&	2.76& 10 &	$9.76\times 10^{13}$&	5&	0.3&	0.27	& 0.17& 0.1&	14&	1.20 \\
S2	&5&	1&	0.23&	4&	0.025&	3.05& 10 &	$3.99\times 10^{13}$&	5&	0.5&	0.45&	0.16 & 0.02&	13&	0.15 \\
S3&	15&	5&	0.84&	20&	0.025&	2.76& 10 &	$3.99\times 10^{12}$&	5&	3&	2.7&	0.17   &0.05&	12&	0.16 \\
S4&	30&	8&	1.88&	20&	0.025&	2.76& 10 &	$9.95\times 10^{11}$&	5&	6&	5.4&	   0.17 &0.08&	28&	0.20 \\
S5&	50&	10&	3.35&	10&	0.025&	3.05& 10 &	$3.95\times 10^{11}$&	5&	10&	9&	0.16 &0.2&	22& 	0.14 \\
W1&	50&	5&	6.67&	48&	0.5&	3.75& 10 &	$4.80\times 10^{11}$&	5&	10&	9.998&	0.15 &18&	96& 	0.14 \\
W2&	50&	2.5&	3.35&	48&	0.5&	3.75& 10 &	$4.80\times 10^{11}$&	5&	10&	9.998& 0.15	&12&	236& 	0.003 \\
W3&	50&	2.5&	3.35&	48&	0.5&	3.75& 10 &	$4.80\times 10^{11}$&	5&	1&	0.998&	0.15 &18&	70&	0.01 \\
W4 &	50&	2.5&	3.35&	5&	0.5&	3.75& 10 &	$4.80\times 10^{11}$&	5&	1&	0.998&	0.15 &10&	178&	0.0002 \\
W5 &	50&	2.5&	3.35&	5&	0.5&	3.75& 10 &	$4.80\times 10^{11}$&	5&	0.5&	0.498& 0.15	&10&	94& 	0.0002 \\
W6&	50&	2.5&	3.35&	5&	0.5&	37.5& 10 &	$4.80\times 10^{12}$&	5&	0.1&	0.098&	0.04 &2.5&	47& 	63.18 \\
W7&	50&	2.5&	3.35&	5&	0.5&	75& 10 &	$9.60\times 10^{12}$&	5&	0.1&	0.098&0.03	&2.5&	47& 	79.80 \\
W8&	50&	2.5&	3.35&	5&	0.5&	150& 10 &	$1.92\times 10^{13}$&	5&	0.1&	0.098& 0.02	&2.5&	47& 	89.10 \\
W9 &	50&	2.5&	3.35&	5&	0.5&	37.5& 50 &	$4.80\times 10^{12}$&	5&	0.1&	0.098&0.04	&1.3&	35& 	0.0002 \\
W10  &	50&	2.5&	3.35&	10&	0.5&	375& 500 &	$4.80\times 10^{13}$&	5&	0.1&	0.098&0.04	&2&	210& 	7.7 \\
W11&	10&	0.5&	0.67&	1&	0.5&	3.75&10&	$1.20\times 10^{13}$&	1&	0.1&	0.098&	0.15 &1&	90&	0.011 \\
\toprule
\end{tabular}
\end{table}

\begin{table}[!h]
\centering
\caption{\label{table:3}The parameters used in CPO simulation. }
\begin{tabular}{c*{16}{c}r}
\toprule
&\multicolumn{12}{c}{Modeling parameters} \\
\cline{2-13}
Case & $R_w$ & $\Delta t $ & & $\Delta R\,(\mu m)$ & & $L_e $ & $N_r\footnote[1]{Number of rays}  $ & $n $ & $\;\;\varphi_0\;\;$ & $L_g $ & $SCTD\footnote[2]{Space charge tube diameter: the diameter of a cylindrical tube in which the desired charge is uniformly deposited.} $ & $E_k\footnote[3]{Kinetic energy} $& Total time & Run time &  $\;\;e^{+}$ lost\\
No. & $(\mu m)$ & $(ps)$  & $\;\;x$ & $y$ & $z\;\;$ & $(mm)$ & & $(cm^{-3})$ & $(V)$ & $(cm)$  & $(\mu m)$& $(eV)$ & $(ns)$ & $(hr)$ & $(\%)$ \\
\hline
S6&	1&	1&	0.05&	0.05&	5&	0.1&	64&	$2.36\times 10^{12}$&	0.007&	0.2&	0.25&	5&	95&	190&	46& \\
S7&	3&	8&	0.1&	0.1&	10&	0.1&	64&	$1.51\times 10^{12}$&	0.04&	1&	0.75&	5&	300&	70&	32& \\
S8&	50&	10&	5&	5&	50&	10&	49&	$2.10\times 10^{10}$&	0.16&	10&	12.5&	5&	430&	24&	9& \\
C1&	50&	0.4&	-&	-&	-&	10&	1&	$4.80\times 10^{10}$&	0.375&	36&	57.7&	5&	700&	20&	-& \\
C2&	50&	0.4&	-&	-&	-&	10&	1&	$1.30\times 10^{11}$&	1&	36&	57.7&	5&	700&	20&	-& \\
C3&	50&	0.4&	-&	-&	-&	10&	1&	$4.80\times 10^{11}$&	3.75&	36&	57.7&	5&	700&	20&	-& \\
\toprule
\end{tabular}
\end{table}
\end{widetext}

Assuming that the radial confinement is maintained by applying high enough magnetic field, a simple analytical density curve is suggested based on the bias voltage required to confine the plasma axially. Calculating the density of the plasma in a microtrap with specific radius, Eq. (\ref{spacecharge}) takes the form
\begin{equation}\label{analyticaleq}
n=\frac{4\pi \epsilon_0 \varphi_0}{q{R_w}^2 (2\ln \frac{R_w}{R_p}+1)},
\end{equation}
in which the density is inversely proportional to ${R_w}^2$   with the fixed  $\varphi_0$ and a constant value of ${R_w}/{R_p}$ . The total density in a microtrap is then recalculated considering its whole volume. For example, if  ${R_w}/{R_p}=\sqrt{3}$ and $\varphi_0$  is  $3.75\,V$, the density in one microtrap will be  $1.6\times 10^{11}\,cm^{-3}$ in the one with $R_w=50\,\mu m$. When the radius of the microtrap is decreased, the analytical density in an individual microtrap increases for a constant space charge potential and constant ratio ${R_w}/{R_p}$. Considering an array of individual microtraps to build the trap, the total number of trapped particles is also inversely proportional to ${R_w}^2$ when the fill factor\footnote[1]{The fraction of total volume of microtraps to the trap volume.}  and the trap volume are constant. Since $n\propto{R_w}^{-2}$ and also $n\propto {\lambda_D}^{-2}$, the ratio of $\lambda_{D}/R_w$ is constant on the curve for the fixed  values of $\varphi_0$, $T$, and ${R_w}/{R_p}$. This ratio should be kept small enough because the equilibrium radial density profile needs an edge with a scale of couple of Debye lengths, as it will be calculated later in this paper. Results of WARP simulation show that the plasma density follows the expected analytical power law with respect to the microtrap radius (i.e. $n\approx k_{WARP} {R_w}^{-2}$), as shown in Fig. \ref{analytical}, suggesting that the Brillouin limit may be surpassed at $1\,\mu m$ microtrap radius while the space charge potential is only  $3.75\,V$. Results from short CPO simulations exhibit an increasing density at a lower rate ($n\approx k_{CPO} {R_w}^{-1}$). If we rewrite the Eq. (\ref{analyticaleq}) as  $n=k{R_w}^{-2}$, the ratios of  $k_{WARP}/k$ and $k_{CPO}/k$  will be equal to $1$ and $0.07$ respectively, showing that the magnitudes of the WARP results are equal to the analytical line while the CPO results pose more than 10 times smaller values and a different power dependence. The CPO results should be considered with caution and will be discussed in section V.

The analytical extrapolation of Eq. (\ref{analyticaleq}) suggests that for a given length of each trap ever smaller trap diameters and larger numbers of parallel traps occupying a fixed volume result in continuously better storage conditions. In the extreme case, replacing a single trap containing $N$ positrons with $N$ traps containing one positron each avoids all plasma complications (e.g. pushing near the Brillouin limit, space charge, etc.) and permits storage times limited only by vacuum conditions. Computer simulations of single component plasmas were initiated to explore this ideal trend and see if the simulation can track an initial density distribution to the equilibrium and if it is possible to store positrons in traps with long aspect ratios for long times. The evolution toward the equilibrium for non-neutral plasma in the PM traps can be so difficult to analyze that one is not able to follow it in detail by analytic theory and so it is important to compute the equilibrium state for a specific trap and plasma. In this study, we evaluated the maximum number of positrons that can be stored in a trap with large aspect ratio and $3$ to $50$ micrometer diameter.

\section{Particle-in-cell (PIC) WARP simulation}
In WARP, the particle-in-cell (PIC) method is employed. A discrete number of real particles Ð positrons in this case Ð are combined to so called macro-particles. The Lorentz equation of motion is employed to advance macro-particles in time. Following each time step, the charge density is calculated via a linear interpolation of the macro-particles position onto a mesh. By solving Poisson's equation, the electrostatic potential is then calculated from the charge density. Currently artificial numerical collisions are included in the WARP simulations to approximate real collisions \cite{Gomberoff}. The rotational symmetry of the microtrap allows the use of the two dimensional version of WARP. It uses a $rz$ field solver with constant potential boundary conditions (i.e. Dirichlet  conditions) at the electrode walls.

\subsection{50 micron radius microtrap}
In this section  the  simulation of the microtrap with $50\,\mu m$ radius is presented. The choice of values for the parameters with the largest influence on numerical noise is discussed. The configuration and dimensions of this microtrap, as well as the magnitudes of magnetic field and electrostatic potentials are consistent with the experimental setup being studied by our research group \cite{Chris}.

\subsubsection{Initial density distribution}
The goal of the modeling is to explore the upper limit in particle density under equilibrium conditions and vanishing loss rates. Before the simulation of this state, the code has to evolve the plasma from its initial as set distribution, which is not an equilibrium state. This time consuming part is minimized if the initial plasma has uniform density and a Boltzmann energy distribution. The initial length of the plasma, as well as the plasma density near the two end electrodes, affected the evolution of the plasma toward the equilibrium. As mentioned earlier, the effective trap potential compels the density to be exponentially small at the end electrodes. Rapid density oscillations occur in the early stages of the simulation when the initial length of the plasma is much shorter than the length of the grounded central tube ($(L_g- L_p \gg 1\,mm$) or the plasma density had a hard edge profile at the ends. Particles located at the ends of the plasma column are accelerated sideways and reflected by the end electrodes. These fluctuations create potential gradients and resulte in heating of the plasma. Subsequently, a higher number of fast particles violate the CFL condition and cause numerical instabilities. Therefore, in the simulations, the plasma was initially $10\,\mu m$ away from each end of the grounded central tube. A "cigar" shape distribution, a built-in function of WARP, was used as the starting configuration of the plasma ends, in which the charge is constant in the center and falls off parabolically at the two end electrodes on a length scale comparable to the radius of the microtrap \cite{WARP}.

As for the initial radial distribution, we stay with the hard edge uniform density (i.e. density falls off sharply to zero). The simulation evolves this into a Òsoft edge distributionÓ where the density of the plasma drops exponentially with radius.

\subsubsection{Time step}
The particle motion is nearly a guiding center motion when the plasma is immersed in a high magnetic field. WARP allows the time step to be larger than the cyclotron period, $\tau$, still correctly calculates the various drifts \cite{Cohen} during the simulation. However, larger time steps result in plasma heating. To minimize heating in the microtrap the time step was calculated from the cyclotron period of

\begin{equation}\label{cyclotronperiod}
\tau =\frac{2\pi m}{q\left|\vec{B}\right|}\approx 5\,ps
\end{equation}

Larger time steps also result in violations of the CFL condition followed by numerical instability and fast radial expansion of the plasma at the earliest stages of simulation (within 100 ps). The simulation of traps with at least one small dimension requires small time steps (especially for higher energy particles) leading to the use of a large number of CPUs on a high performance computer and weeks of running time.

\subsubsection{Mesh size}

\begin{figure}[!h]
\includegraphics[width=75mm]{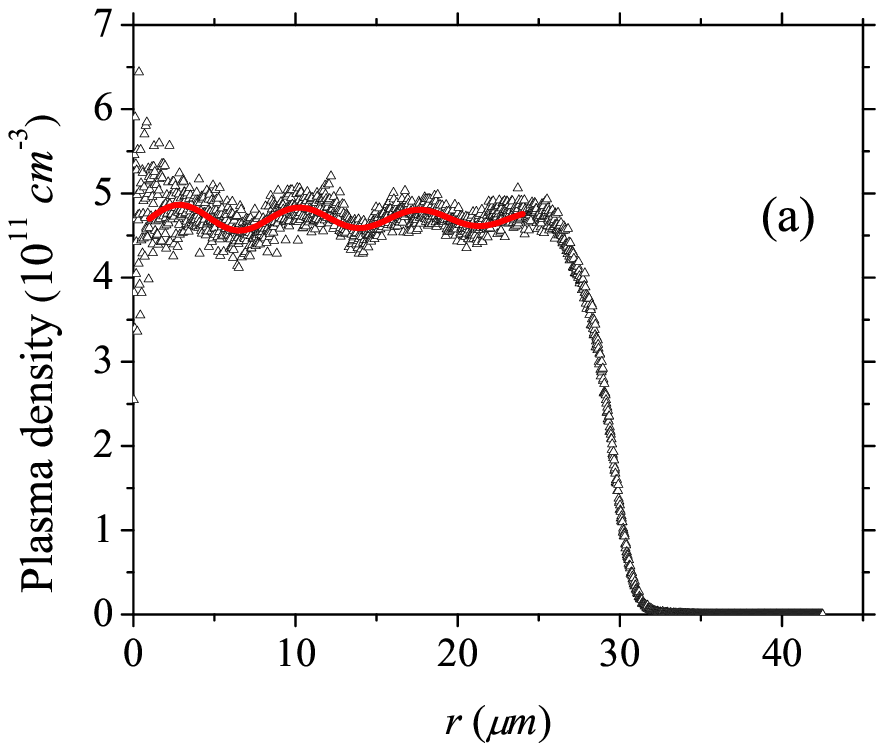}
\includegraphics[width=75mm]{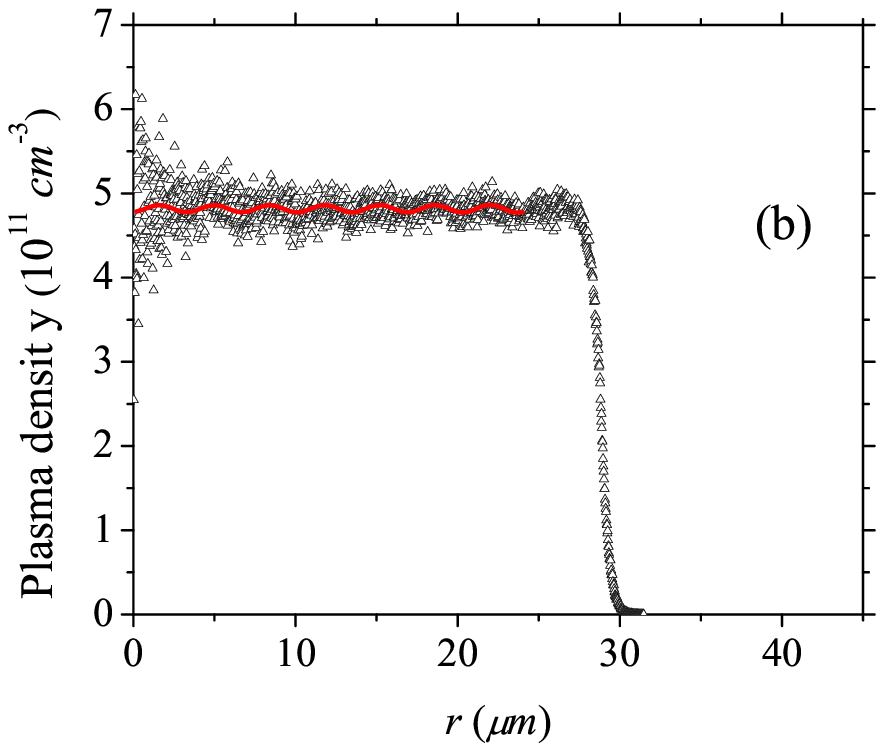}
\caption{\label{meshwave}The density histogram with  $0.02\,\mu m$ bins at $t=3.2\,\mu s$, $\vartriangle$. Fit to data with Eq. (\ref{waveequation}), {\color{red}$-$}, follows the wavelike behavior for (a) the case W1 and (b) the case W2. No clear ordered wave is seen for (c) the case W4 with smaller positron weight.}
\end{figure}

\begin{figure}[!h]
\includegraphics[width=75mm]{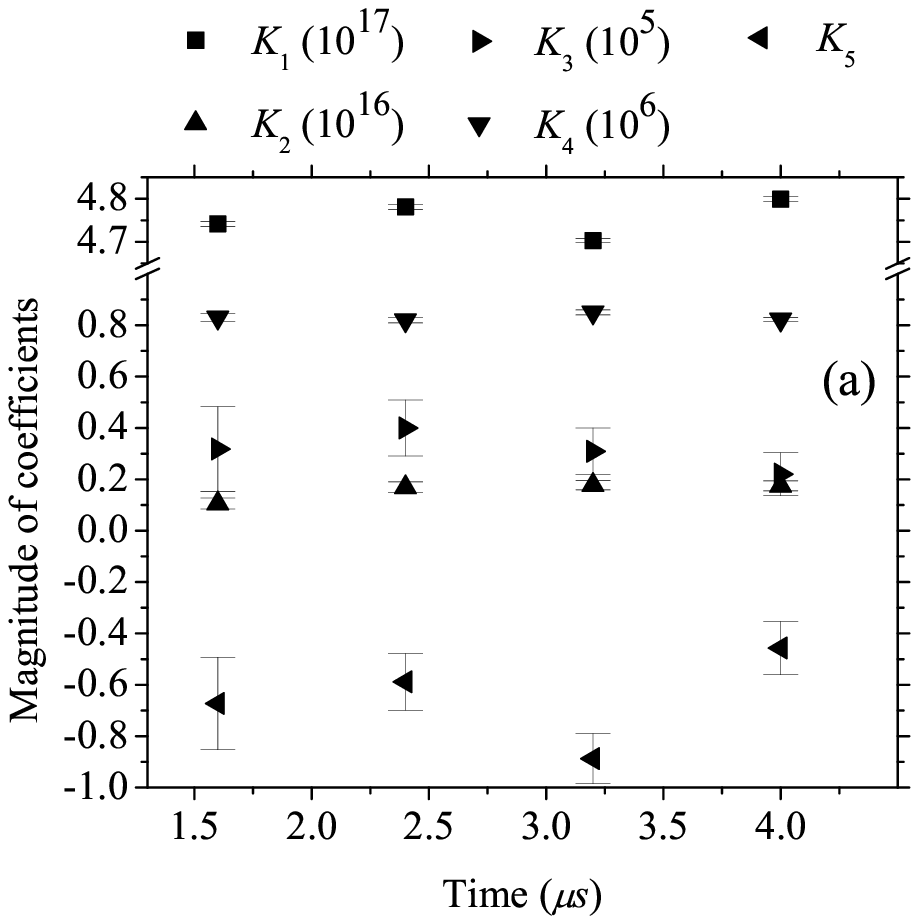}
\includegraphics[width=72mm]{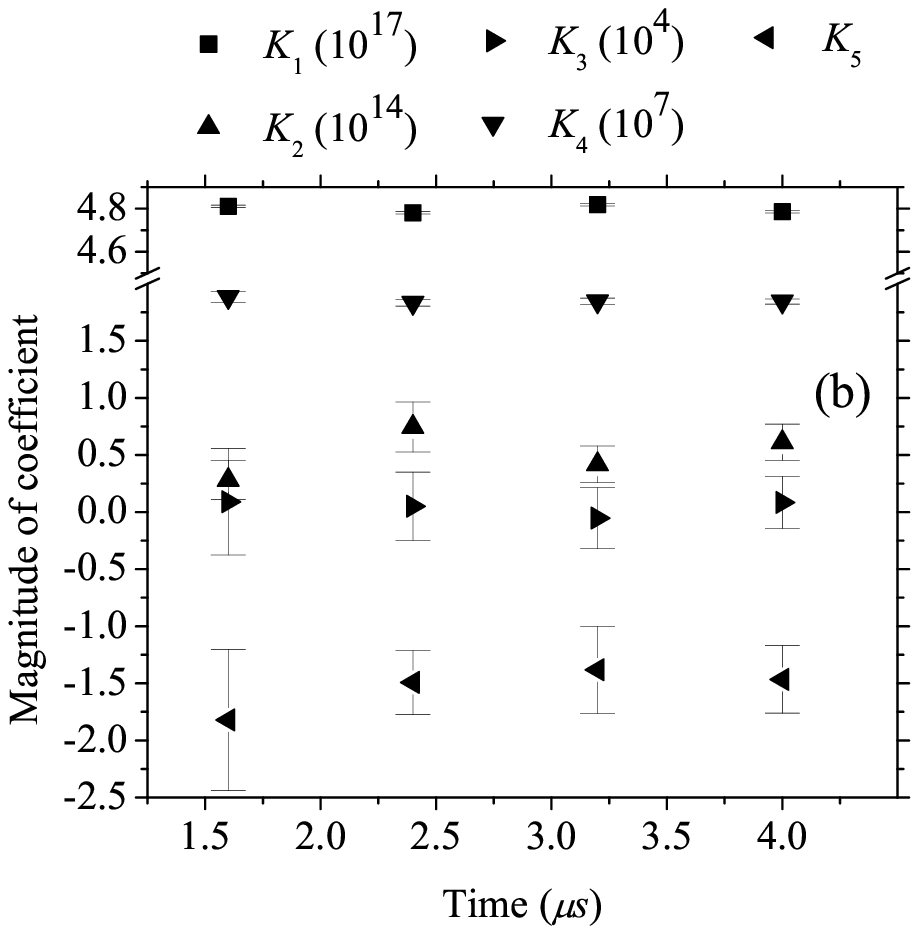}
\caption{\label{coefficient}The magnitude of coefficients variation in time from the fitting Eq. (\ref{waveequation}) for (a) the case W1 and (b) the case W2. $k_1$  is the mean density, $\blacksquare$, $k_2$  is the amplitude, $\blacktriangle$, $k_3$  is the decay rate with radius, $\blacktriangleright$, $k_4$  is the frequency in radial units, $\blacktriangledown$, and $k_5$  is a phase shift, $\blacktriangleleft$.}
\end{figure}

Simulations were carried out on discrete square mesh cells in the $rz$ plane. The mesh size can't be chosen larger than the Debye length since the Debye length would not be resolved in simulation. This adds to the difficulty of simulating high density plasmas. To illustrate the effect of mesh size on the simulation, two mesh configurations were compared in the cases W1 and W2, whose parameters are listed in Table II. In the case of W1 the mesh size was equal to $6.7\,\mu m$ for $r$ (radial) and $z$ (parallel to the magnetic field and the trap axis). The time step was $5\,ps$. After  $1.6\,\mu s$, wavelike variations develop in the radial density profile. These Òdensity wavesÓ remain constant similar to a standing wave. This is illustrated in Fig. \ref{meshwave}(a) at  $t=3.2\,\mu s$. To investigate the behaviors of these waves, the density histograms are fitted with equation
\begin{equation}\label{waveequation}
n=k_1+k_2\exp (-k_3 r) \sin (k_4 r+k_5),
\end{equation}
 where $k_1$ is the mean density, $k_2$  is the amplitude,  $k_3$ is the decay rate with radius, $k_4$  is the frequency in radial units, and $k_5$  is a phase shift. These coefficients are plotted as a function of time in Fig. \ref{coefficient}(a). The magnitude of  $k_4$ remain constant (varies $\pm2\%$) in time. The wave has a wavelength of  $7.52\pm0.15\,\mu m$, close to the mesh size in the case W1, and so may suggest that the fluctuation of the radial density distribution inside the plasma may be due to the mesh size. To investigate the validity of this claim, the mesh size was reduced to  $3.35\,\mu m$ in the case W2. The time step was reduced to $2.5\,ps$ in order to avoid violating the CFL conditions  $\Delta R/\Delta t=1.34\times 10^{6}\,ms^{-1}$. The "density wave" inside the plasma exhibited the shorter wavelength of  $3.38\pm0.005\,\mu m$, which is in excellent accord with the smaller mesh size of  $3.35\,\mu m$ in the case W2. The magnitude of  $k_4$ varied $1.5\%$ as shown in Fig. \ref{coefficient}(b). The fixed value of  $k_5$ (did not change effectively in both cases) along with the constant  $k_4$ imply that the waves were fixed in radial position ($\pm 0.25\,\mu m$) and so support the idea of the waves dependency on the mesh size.

\begin{figure}[!h]
\includegraphics[width=75mm]{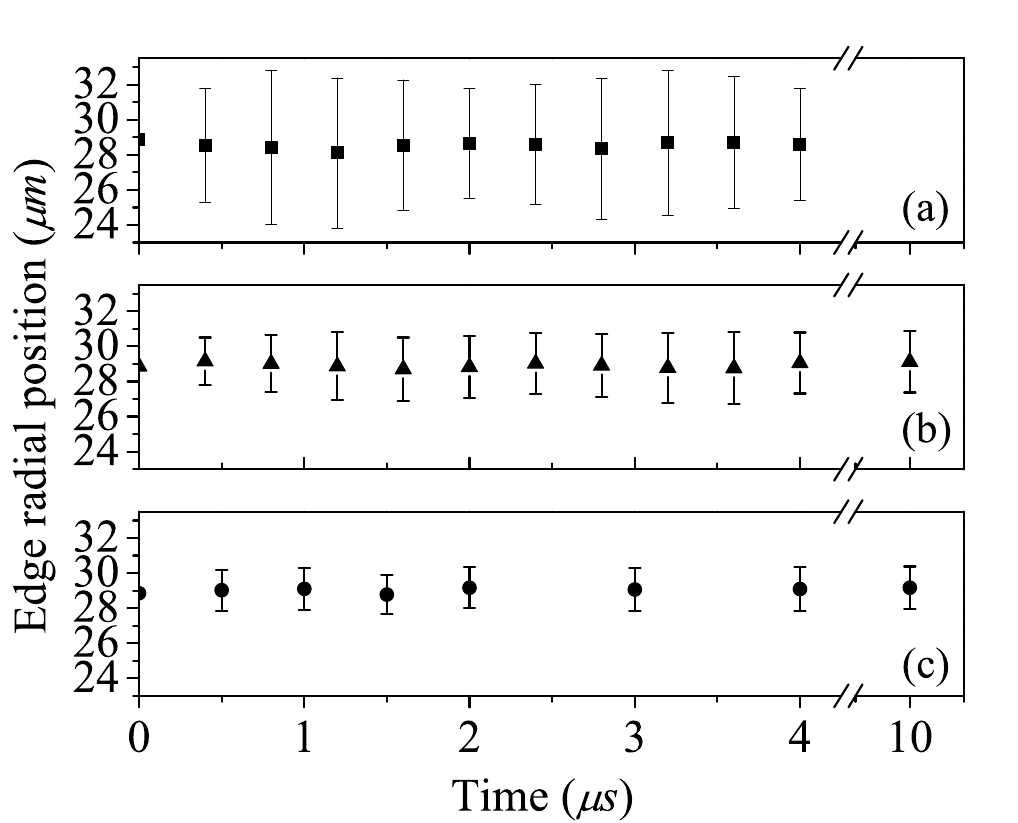}
\caption{\label{edge}The mean radial position of the edge as a function of time for the case W1, $\blacksquare$, and the case W2, $\blacktriangle$, and the case W4, $\bullet$. The edge width is shown as error bars.}
\end{figure}

The amplitude of the waves dropped by an order of magnitude when the mesh size was cut in half from $6.7$  to $3.35\,\mu m$. Furthermore, the density waves which showed larger amplitude was decayed by an order of magnitude faster with respect to the radius. Also, the mean density in the wave region oscillated three times larger in the case of the larger mesh size. This oscillation accompanied by the radial oscillation of the plasma, which was seen by studying the plasma edge behavior in Fig. \ref{edge}. The radius and width of the plasma edge were oscillating much larger in the case of the larger mesh size. This case experienced a softer plasma edge and the density profile smeared out further in the same time scale.

\begin{figure}[!h]
\includegraphics[width=75mm]{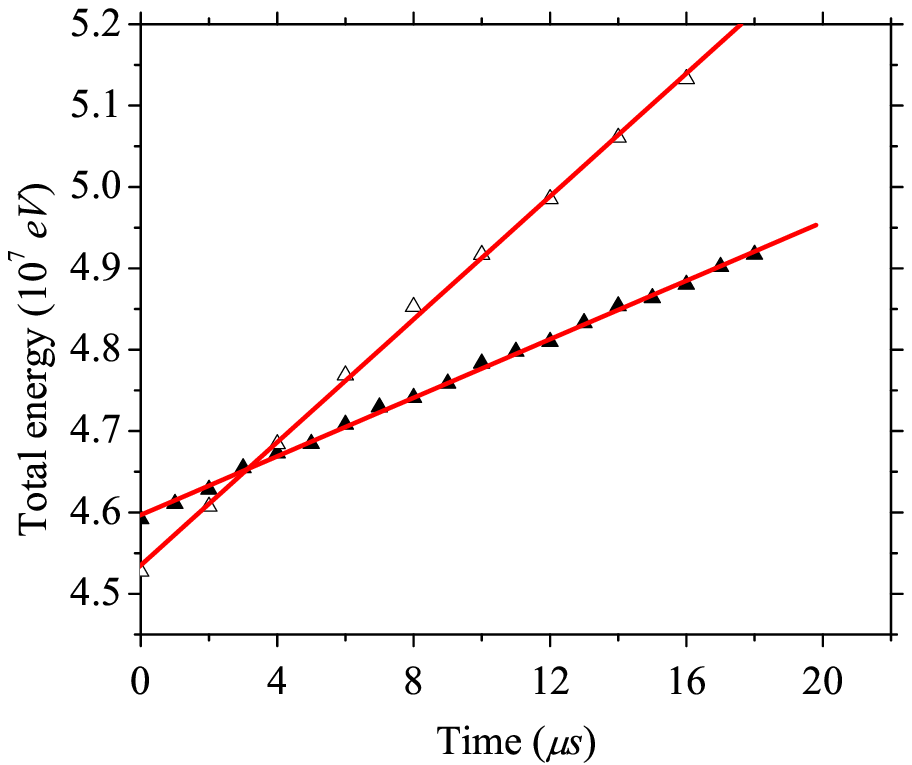}
\caption{\label{TotalEnergyW1W2}The total energy of the plasma is plotted in time, which helped to evaluate the degree of numerical heating, for the case W1 with mesh size of $6.7\,\mu m$, $\blacktriangle$, and for the case W2 with mesh size of $3.35\,\mu m$, $\vartriangle$, which exhibits linear growth with slopes of $3.7\times 10^8$ and $1.8\times 10^8\,eVs^{-1}$, respectively.}
\end{figure}

Fig. \ref{TotalEnergyW1W2} shows the total energy over time in the case W1 and W2. Because of the overall non-conservation of the total energy due to the mesh, the numerical heating is unavoidable and there is no constraint in Warp that keeps the energy conserved. However, the degree of numerical heating decreased in half when the mesh size was cut in half. It also helped reduce both wavelength and amplitude of the density waves and decreased the plasma oscillation but could not avoid it completely. The key to solve this problem relies on providing enough number of particles per mesh cell.

\subsubsection{Positron weight}
WARP is based on the PIC model and does not track individual particles, such as positrons; a selectable number is combined into macro-particles. The positron weight is the number of real particles that each simulation macro-particle represents. Larger value of the positron weight reduces the computation time, while a smaller number helps the plasma to evolve toward computational equilibrium. In the PIC model, the short-range forces are not correctly modeled, with the shortest ranges being of order of the mesh cell size \cite{Birdsall}. This approximation and the weight of the particles $>1$, give rise to artificially large collision cross sections and lead to the heating. Computational equilibrium in conventional traps has been demonstrated previously \cite{Gomberoff} by WARP where the initial number of macro-particles per cell was of the order of unity or less. This helped reduce the typical axial oscillations due to the high level of numerical collisions.

\begin{figure}[!h]
\includegraphics[width=75mm]{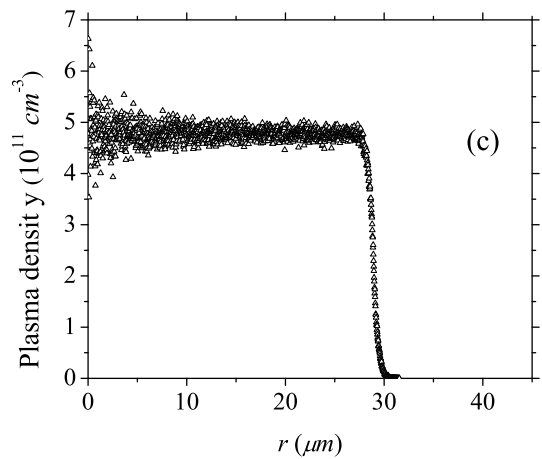}
\caption{\label{meshwaveW4}The density histogram with  $0.02\,\mu m$ bins at $t=3.2\,\mu s$, $\vartriangle$, for the case W4.}
\end{figure}

\begin{figure}[!h]
\includegraphics[width=75mm]{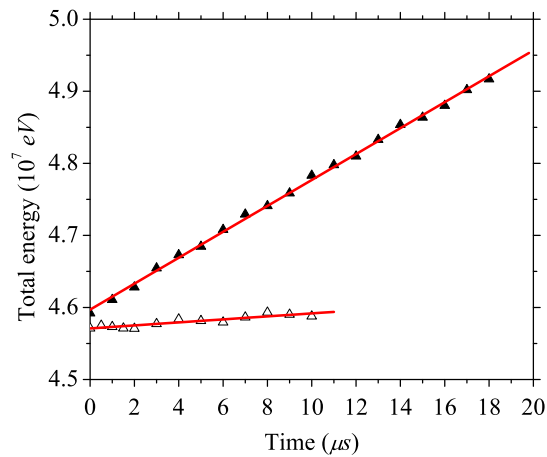}
\caption{\label{TotalEnergy}The total energy of the plasma is plotted in time, which helped to evaluate the degree of numerical heating, for the case W3 with positron weight of 48, $\blacktriangle$, and for the case W4 with positron weight equal to 5, $\vartriangle$, which exhibits linear growth with slopes of $1.8\times 10^5$ and $2\times 10^4\,eVs^{-1}$, respectively.}
\end{figure}

For the first cases in our simulation (W1, W2, and W3) the positron weight had larger values so the initial number of macro-particles per mesh cell was about or less than unity. The case W4 was run with the positron weight of about ten times smaller than the case W3 while all other parameters were consistent. These cases were modeled for the trap with $1\,cm$ length grounded central tube to avoid very high computational times as if we would keep the original length of 10 cm, it would cost couple of months computing on the same computational hardware on the high performance computer. As illustrated in Fig. \ref{meshwaveW4}, no clear density wave was experienced in the case of smaller positron weight (i.e. $PW=5$), the case W4. Simulations using smaller positron weight (i.e. higher number of particles per mesh cell), experience a lower degree of numerical heating in the micro-scale plasma simulation, as shown in Fig. \ref{TotalEnergy}. As stated before, the numerical heating is unavoidable due to the mesh and it would not be possible to avoid the noise completely. However, the degree of heating can be reduced so that the time scale in which the instability dominates becomes long compared to the time required for the charges to attain the thermal equilibrium.

\subsubsection{Toward the equilibrium}

\begin{figure}[!h]
\includegraphics[width=75mm]{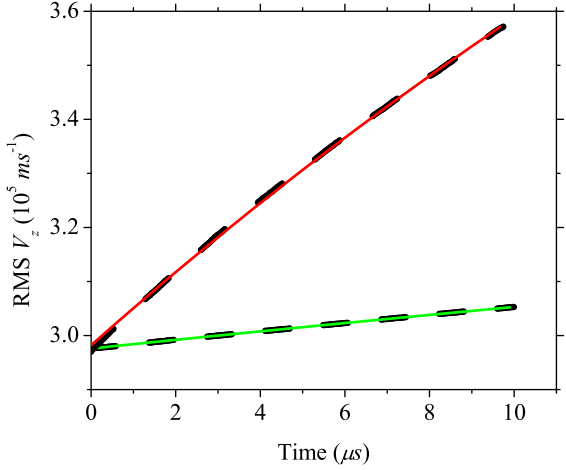}
\caption{\label{RMSvz}The evolution of RMS $V_z$, dashed black lines, for the cases W3 and W4. The fit to data with RMS $V_z=c_1\exp(-t_1/c_2)+c_3$ calculates $c_1=-248760\pm100$, $c_2=4\times10^{-5}\pm1\times10^{-8}$, and $c_3=546968\pm103$ for the case W3, {\color{red}$-$}, and calculates $c_1=-74327\pm132$, $c_2=9.15\times10^{-5}\pm1\times10^{-7}$, and $c_3=371920\pm134$ for the case W4, {\color{green}$-$}.}
\end{figure}

Typically, there are two stages of equilibrium for a plasma. The collisions play the main role to bring the plasma to the local equilibrium. In this stage the rotating frequency and the temperature is dependent of $r$ while these values become constant on very long time scales (few seconds) in the case of global equilibrium with the help of shear forces and radial heat transport \cite{Dubin}. As the plasma attains equilibrium in which there is no transport across the magnetic field, the root mean square of axial velocity (RMS $V_z$) should reach to a constant value, although it has been shown \cite{Gomberoff} that there could be some oscillations in the this value when a plasma nears the equilibrium in a PM trap. The evolution of RMS $V_z$ is plotted in Fig. \ref{RMSvz} for the cases W3 and W4, showing that the values increase with a decaying rate. The curves are fitted with the exponential decay prediction of theory \cite{Ichimaru} to give the relaxation rate. The initial value of $2.97\times 10^5\,ms^{-1}$ would increase to $\sim 5.46\times10^5$ and $\sim 3.71\times10^5\,ms^{-1}$ at longer times with the decay rate of $2.5\times 10^4$ and $1.1\times 10^4$ (i.e. ${c_2}^{-1}$) with a half-life of $17.1$ and $33.9\,\mu s$, respectively.

\begin{figure}[!h]
\includegraphics[width=75mm]{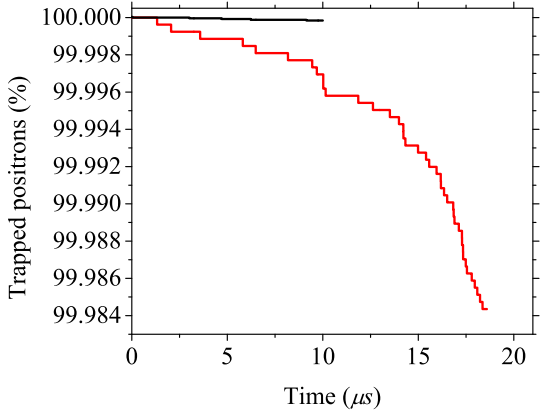}
\caption{\label{trapped}The time histories of the number of trapped particles in the case W3, {\color{red}$-$}, and the case W4, $-$.}
\end{figure}

\begin{figure}[!h]
\includegraphics[width=75mm]{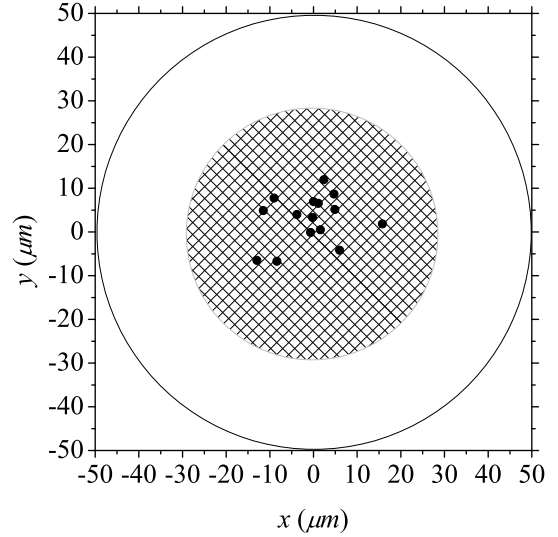}
\caption{\label{lostposition}The cross section of one end electrode shows the position of the lost particles in the case W3, $\bullet$. The patterned surface represents the initial plasma.}
\end{figure}

The rapid increase of RMS $V_z$ in the case W3 was due to the high degree of numerical heating. Before the plasma attains equilibrium, this instability causes overheating a lot of particles which are not longer confined by the end electrodes. While no particles were lost by reaching to the cylinder wall, about $3.0\times10^{-3}\%$ of particles escaped axially across the $10\,V$ end electrodes after $10\,\mu s$, as shown in Fig. \ref{trapped}, in the case W3 while the loss rate was increasing. Figure \ref{lostposition} illustrates that most of these particles were lost close to the central axis of the trap where the space charge potential is maximum. So the loss was not due to outward drift. Simulation result for the case W4 show that only $1.6\times10^{-4}\%$ of particles (i.e. four macro-particles) were lost after $10\,\mu s$, implied that the numerical instabilities were unlikely to grow and dominate at larger timescales.

{Because the implemented initial axial distribution in the simulation was just an estimate of the equilibrium distribution (e.g. we did not actually solve where the density exactly drops to zero near the end electrodes while the plasma is at rest), the axial velocity distribution deviated slightly in time from initial Maxwellian. But as the plasma reaches the first stage of equilibrium (local equilibrium), the velocity distribution became Maxwellian again. The longitudinal force balance was satisfied along each magnetic field lines,

\begin{figure}[!h]
\includegraphics[width=75mm]{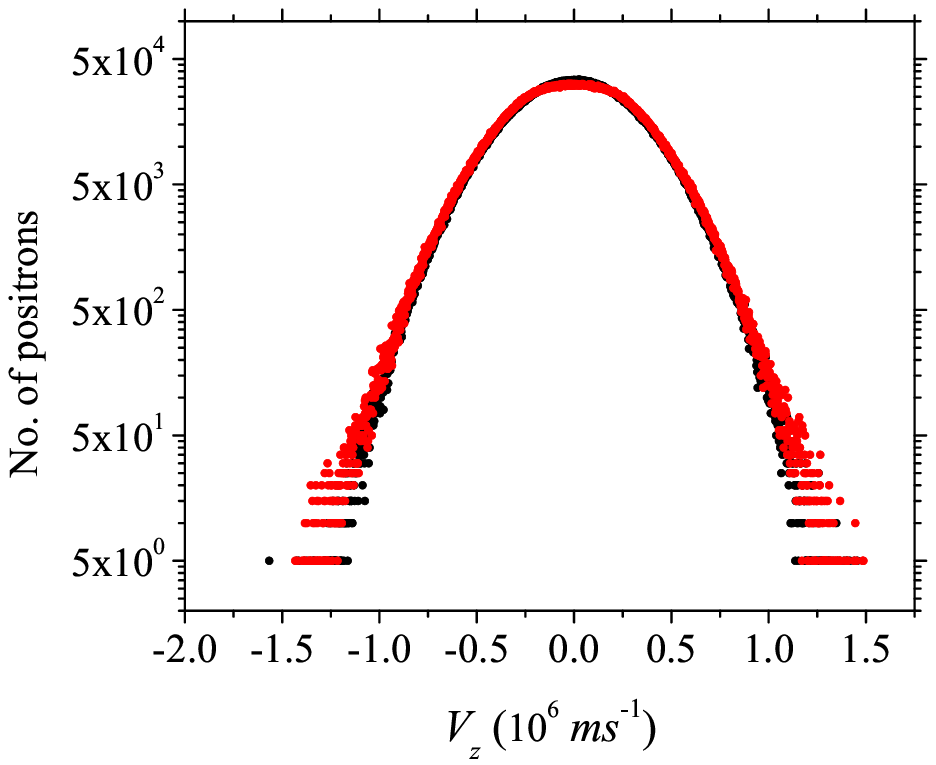}
\includegraphics[width=75mm]{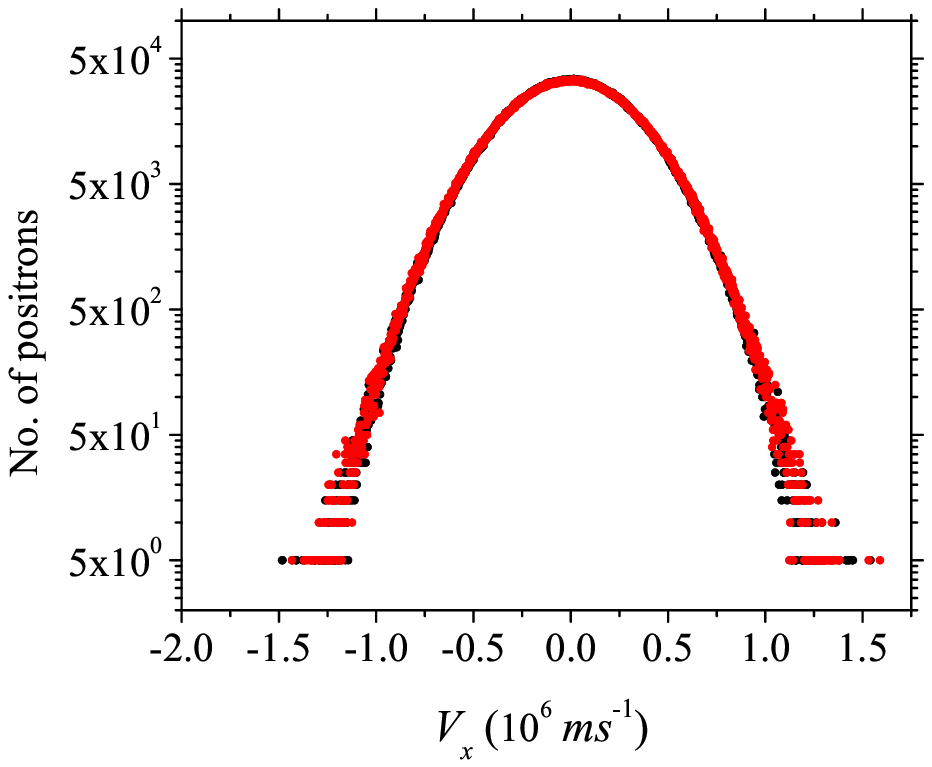}
\caption{\label{Vzx}{\color{blue} $V_z$ and (b) $V_x$ histograms with $2\times10^3\,ms^{-1}$ bins at $t = 0$, $\bullet$, and at $t = 10\,\mu s$, {\color{red}$\bullet$}, in the case W4.}}
\end{figure}

\begin{equation}\label{forcebalancez}
nqE_z+\partial p/ \partial z=0.
\end{equation}
The solution to this equation at each cylindrical shell with negligible thickness and the z axis symmetry is a Blotzmann factor \cite{Dubin},
\begin{equation}\label{Blotzmannfactor}
n(r,z,t)/N(r,t)= \frac{\exp [-q\phi(r,z,t)/kT(r,t)]}{\int\limits_{-\infty}^{\infty}{\exp [-q\phi(r,z,t)/kT(r,t)]dz}},
\end{equation}
where $N$ is the normalized density for each shell and $\phi$ is the potential on the laboratory frame. Fig. \ref{PhiZ} shows the $\phi -z$ phase for simulation particles at each cylindrical shell. Since the density is uniform at each cylindrical shell far from the end electrodes ($i.e.$ the solution exists there), it is more of interest to study the distribution near the end electrodes where the potential rises. The distribution at $t = 10\,\mu s$ showed a good conformity with the analytical solution in the case W4, as shown in Fig. \ref{nN}, which proved the existence of the local equilibrium. The potentials expanded the plasma axially on a time scale much shorter than the final equilibrium time yet much longer than the axial bounce time.

\begin{figure}[!h]
\includegraphics[width=75mm]{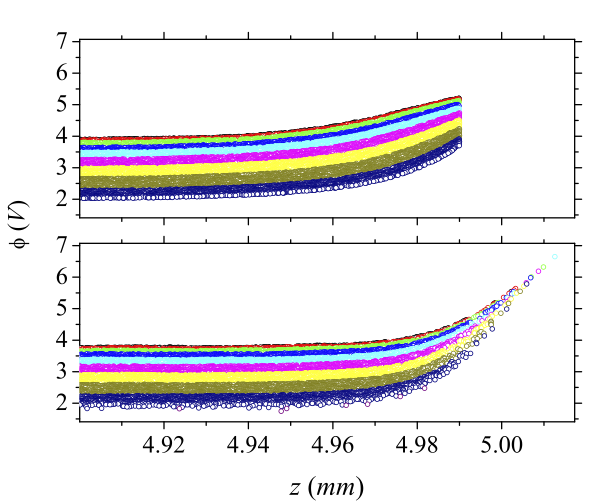}
\caption{\label{PhiZ} {$\phi -z$ phase for simulation particles at each cylindrical shell at $t = 0\,\mu s$, top figure, and $t = 10\,\mu s$, bottom one, in the case W4, shown on the region where the main tube meets the end electrode. Shells thicknesses are equal to the mesh size, $3.35\,\mu m$.}}
\end{figure}

\begin{figure}[!h]
\includegraphics[width=85mm]{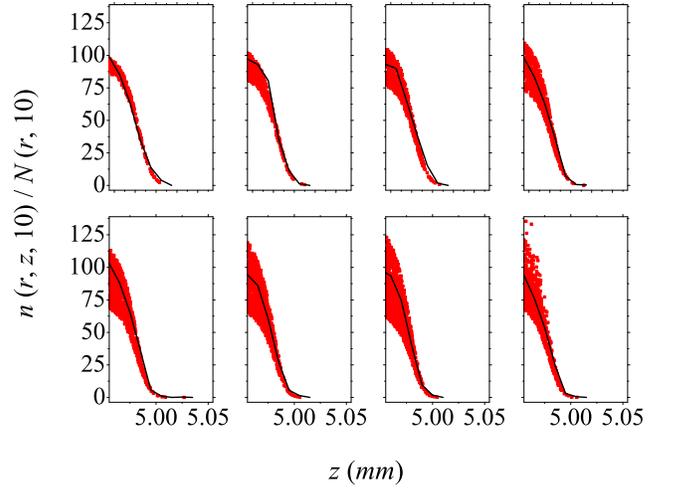}
\caption{\label{nN} {The curve of $n(r,z,t)/N(r,t)$, $-$,  at each cylindrical shell, and the right side value of Eq. (\ref{Blotzmannfactor}) for each particle, {\color{red}$\bullet$}, at  $t = 10\,\mu s$ in the case W4. Data are shown on the region where the main tube meets the end electrode. Shells thicknesses are equal to the mesh size, $3.35\,\mu m$.}}
\end{figure}

The plasma temperature was radially uniform initially at $0.5\,eV$, which remained uniform throughout the evolution at the case W4, rising in value to  $0.5190\,eV$ by $t=  10\,\mu s$ as observed in Fig. \ref{Temperature}. Data are fitted with the Boltzmann energy distribution function, Eq. (\ref{Boltzmanenergy}). The plasma showed good azimuthal symmetry in density and temperature. The temperature and its perpendicular ,$T_\bot$, and longitudinal, $T_{||}$, values are illustrated versus radius at $t=  10\,\mu s$ in the Fig. \ref{Temperature2}. The values are monotonic throughout the radius except at the plasma's edge. The initial hard edge profile imposed a high pressure force on the positrons at the edge and smeared them out in the early times of the simulation, creating a relaxed soft edge with $T_\bot$ changed and $T_{||}$ unchanged and an anisotropic temperature distribution. Future work can include changing the initial density distribution to an exponentially decreasing one for large radii of plasma in order to avoid this early stage heating. This physical heating will affect the density distribution at the edge as it will be described later.

\begin{equation}\label{Boltzmanenergy}
f_E=2\sqrt{\frac{E_k}{\pi}}(\frac{1}{T(eV)})^{3/2}\exp(-\frac{E_k}{T(eV)}).
\end{equation}

\begin{figure}[!h]
\includegraphics[width=75mm]{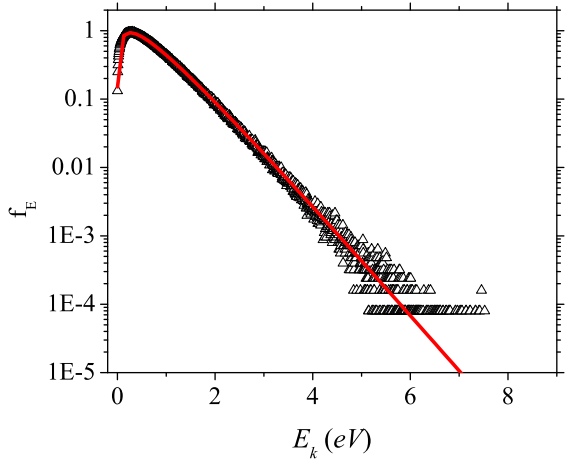}
\caption{\label{Temperature}Kinetic energy histogram at $t=10\,\mu s$ with $5\times 10^{-3}\,eV$ bins for the case W4. Fitted data with Eq. (\ref{Boltzmanenergy}) calculates $T=0.5190\pm2\times10^{-4}$.}
\end{figure}

\begin{figure}[!h]
\includegraphics[width=75mm]{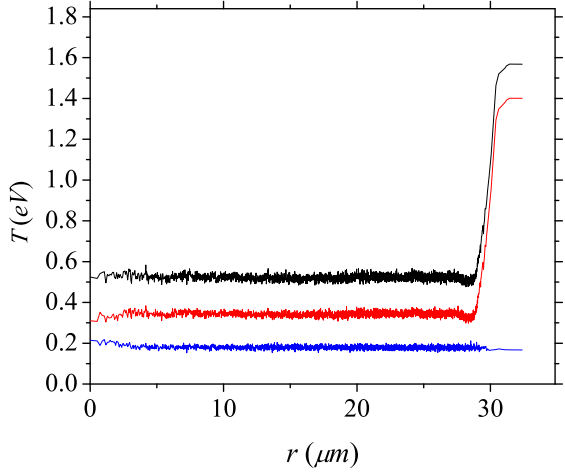}
\caption{\label{Temperature2}$T$, $-$, $T_\bot$, {\color{red}$-$}, and $T_{||}$, {\color{blue}$-$}, vs. plasma radius at $t=10\,\mu s$ in the case W4.}
\end{figure}

Radial variation in the rotation frequency  in Fig. \ref{omega} shows that the flow had substantial shear at the time of $10 \,\mu s$, rotating faster at outer region compared to the center. The soft edge was rotating even much faster because of the early edge heating as previously described. The inner section ($r<13 \,\mu m$) exhibits the frequency of $8.95\times 10^{8}\, s^{-1}$ which is about 50\% higher than that calculated from Eq. (\ref{rotationfrequency2}), $6.2\times 10^8$. The rotation is also quite uniform in $z$ as expected in local equilibrium, ${\partial \omega}/{\partial z}=0$. The density profile has evolved to a form that one expects for thermal equilibrium but it takes few seconds for the shear in the rotational flow to vanish as reported in the experiments. However, local equilibrium states have been reported to be observed at times of the order of mili-seconds \cite{Hyatt}. This involves turbulent flows and large density fluctuations after the injection while the initial state of plasma in our simulation was not that far away from the equilibrium and that's why the local equilibrium was reached in comparatively shorter time scales. Moreover, we have a quite narrower plasma which is expected to attain the equilibrium faster. In another experiment, the plasma was reported to be in local equilibrium for $t\approx {V_{th}}^{-1}$ \cite{Hyatt} and a rotation frequency profile very similar to what was shown in Fig. \ref{omega} was seen experimentally before the global equilibrium was attained \cite{DriscollMalmbergFine}.

\begin{figure}[!h]
\includegraphics[width=75mm]{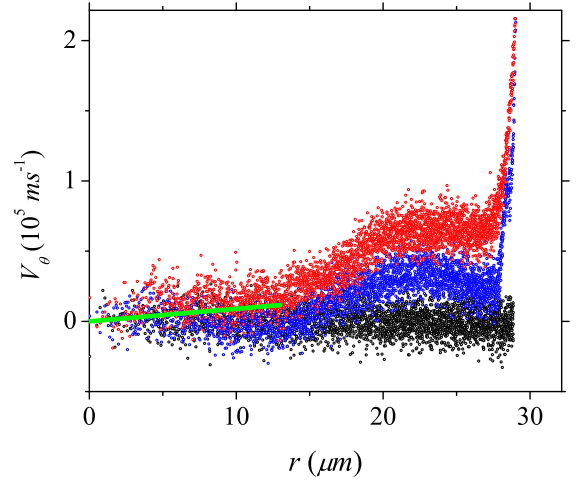}
\caption{\label{omega} Azimuthal velocity as a function of the plasma radius at $t = 0$, $\bullet$, $1$, {\color{blue}$\bullet$}, and $10\,\mu s$, {\color{red}$\bullet$}, in the case W4. The inner section ($r<13 \,\mu m$) exhibits the rotation frequency of $8.95\times 10^{8}\, s^{-1}$ at $t = 10\,\mu s$.}
\end{figure}

Expecting that the force balance is satisfied in radial direction by the equilibrium distribution, we can write
\begin{equation}\label{forcebalance}
nqV_\theta B-nm{V_\theta}^2/r=nqE_r-\partial p/\partial r,
\end{equation}
where $p=nkT$ is the pressure and $E_r=q\int\limits_{0}^{r}nrdr/r\epsilon_0$ for a symmetric plasma. Dividing both sides of Eq. (\ref{forcebalance}) to $n/r$ and differentiating, we can write
\begin{equation}\label{Matheq}
\frac{q^2}{\epsilon_0}nr-kT(-\frac{\partial n}{\partial r}+\frac{1}{r} n\frac{\partial n}{\partial r}+n\frac{\partial^2 n}{\partial r^2})-2r(q\omega_a B-m{\omega_a}^2)=0.
\end{equation}
The parameter $\alpha$ is defined as
\begin{equation}\label{alpha}
\alpha=1-\frac{\omega_a}{\omega},
\end{equation}
where $\omega$ is obtained from Eq. (\ref{rotationfrequency2}). Solutions to Eq. (\ref{Matheq}) shown in Fig. \ref{Math} are derived by numerically integrating for different values of $\alpha$. Confinement (i.e. $n\rightarrow0$ as $r\rightarrow\infty  $) requires that $\alpha >0$. Dividing the $r$ axis by $\lambda_D$ and the $n$ axis by $n_{r=0}$, we got a unique curves regardless of density and temperature.

\begin{figure}[!h]
\includegraphics[width=75mm]{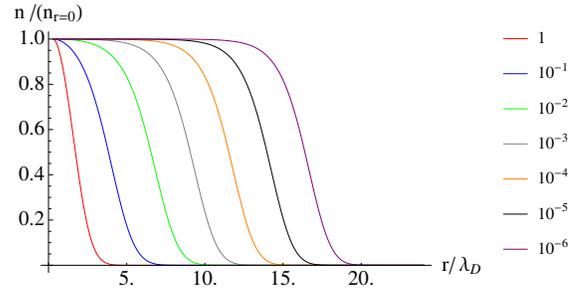}
\caption{\label{Math} Solutions for radial density distribution in Eq. (\ref{forcebalance}) for different values of $\alpha$.}
\end{figure}

Therefore, the plasma density is nearly constant out at the equilibrium to some radius and there drops to zero on the scale of a few Debye lengths. This statement can be understood by the fact that positrons arrange their positions at the state of equilibrium so that any externally imposed electrostatic field is Debye shielded out \cite{Davidsonbook1990}. Similar solutions of equilibrium have been given \cite{ONeilDriscoll} by solving the Boltzmann distribution while total Hamiltonian and total angular momentum are conserved.

It is assumed in these calculations that the temperature is monotonic throughout the radius. Now if the plasma poses a warmer edge, similar to what we got in the simulation, the density drops to zero in a comparatively shorter length, as shown in Fig. \ref{Density}(c). Note that  the edge width, $\approx 2.5\,\mu m$, is even much smaller than the Debye length, $\approx 7.5\,\mu m$. While the edge showed a stable width in the time frame of our simulation, as shown in Fig. \ref{edge}, it would broaden in a very longer time scales, typically couple of seconds, by the help of heat transport.

The total particle energy can be written as the Hamiltonian, which is invariant in time. It includes the kinetic energy, electrostatic energy, electrostatic interaction energy of the charges with each other, and electrostatic interaction energy of the charges with their images. The images charges are included implicitly in the Poisson solve boundary condition of WARP. With the Dirichlet boundary condition, the potential on the boundary is fixed. It's as if there were image charges present and they go to the locations needed to get the potential to its fixed value. In a real metal boundary, the free electrons would move around so that the transverse electric fields on the metal surface vanish. The code doesn't calculate the locations of these charges, but they are implied by fixing the value of the potential on the surface.

Angular momentum can be written as \cite{Oneil}
\begin{equation}\label{angularmomentum2}
P_\theta=\sum\limits_{i=1}^{N} {mv_{{\theta}_{i}} r_i+qB{r_i}^2/{2c}},
\end{equation}
in which the kinetic part is very smaller and can be ignored in existence of a large enough magnetic field and low enough densities (i.e. $\omega_p\ll \omega_c$). It implies that the confinement is guaranteed if there is a constraint on the mean square radius of the plasma,
\begin{equation}\label{RMSr}
\sum\limits_{i=1}^{N} {r_i}^2\simeq const.
\end{equation}
However, the plasma can expand in large time scales due to the asymmetries and collisions with neutrals which change this value. Mean square radius of the plasma in the case W4 exhibited no clear expansion after $10\,\mu s$, which confirms that the angular momentum is conserved. All of the plasma dynamics involve only internal interactions in our simulation, so conservation of angular momentum (i.e. constant mean square radius) means that in case of an expansion for very larger time scales, only small fraction of particles can move from radii smaller than $\approx R_w/\sqrt 3$ to the radius of $R_w$ and the others must remain confined.

\subsubsection{Length-dependent relaxation}

A strong enhancement has been showed in cross-field collisional heat transport due to the long-range collisions compared to the classical theory \cite{DubinTransport}. Particles on field lines separated by up to $\lambda_D$ can exchange axial velocities in this long-range theory (for plasmas with $\lambda_D > r_c$ where $r_c$ is the cyclotron radius) and so the heat is transported independent of density and magnetic field, scaling only with temperature as $T^{-1/2}$. In the case W5, the influence of the length of the trap on the plasma relaxation was examined by dividing the trap length in half to the case W4. The numerical heating on both cases are very small and comparable, as shown in Fig. \ref{TotalEW4W5}, and so we can neglect the effect of simulation noises in comparison. Figure \ref{RMSvz2} are data for root mean square of axial velocities for the cases W4 and W5. It exhibits comparable values of final RMS $V_z$ ($c_3$) and relaxation rate ($1/c_2$) in two cases, suggesting that the heat transport is also independent of the plasma length.

\begin{figure}[!h]
\includegraphics[width=75mm]{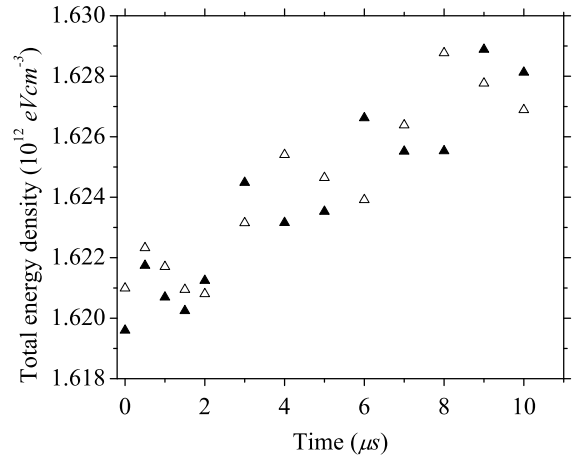}
\caption{\label{TotalEW4W5}Comparison of the increase of the plasma total energy density in time between the case W4 with $1\,cm$ trap length, $\vartriangle$, and the case W5 with the trap length equal to $5\,mm$, $\blacktriangle$, exhibiting similar linear trend.}
\end{figure}

\begin{figure}[!h]
\includegraphics[width=75mm]{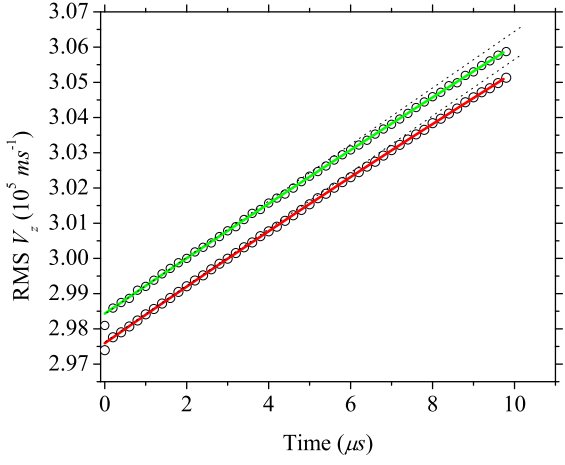}
\caption{\label{RMSvz2}The evolution of RMS $V_z$, $\circ$, in time for the cases W4 and W5. The fit to data with RMS $V_z=c_1\exp(-t/c_2)+c_3$ calculates $c_1=-74327\pm132$, $c_2=9.15\times10^{-5}\pm1\times10^{-7}$, and $c_3=371920\pm134$ for the case W4, {\color{red}$-$}, and calculates $c_1=-94041\pm370$, $c_2=1.18\times10^{-4}\pm4\times10^{-7}$, and $c_3=392472\pm370$ for the case W5, {\color{green}$-$}.}
\end{figure}

Plasmas with different lengths in the cases W4 and W5 experienced a similar temperature evolution during $10\, \mu s$ of simulation time and can be observed in Figs. \ref{T1cm} and \ref{T1cm5mm}. The temperature was recalculated by ruling out the effect of the numerical heating. It is implied that both would reach to the same final temperature at longer simulation times and length of the plasma column does not also affect on its thermalization.

\begin{figure}[!h]
\includegraphics[width=75mm]{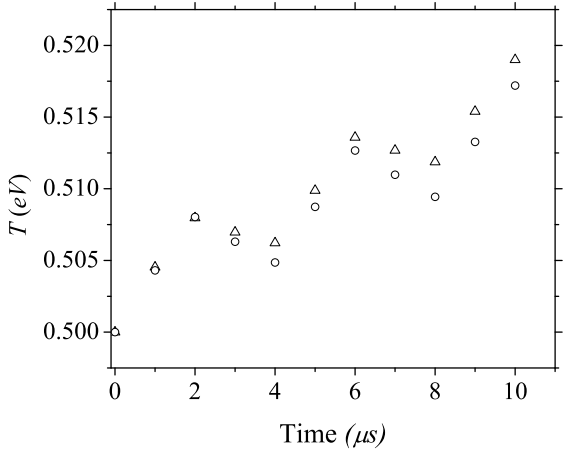}
\caption{\label{T1cm}The plasma temperature increase in time for the case W4, $\vartriangle$. The temperature was recalculated, $\circ$, by ruling out the heating due to the simulation noise.}
\end{figure}

\begin{figure}[!h]
\includegraphics[width=75mm]{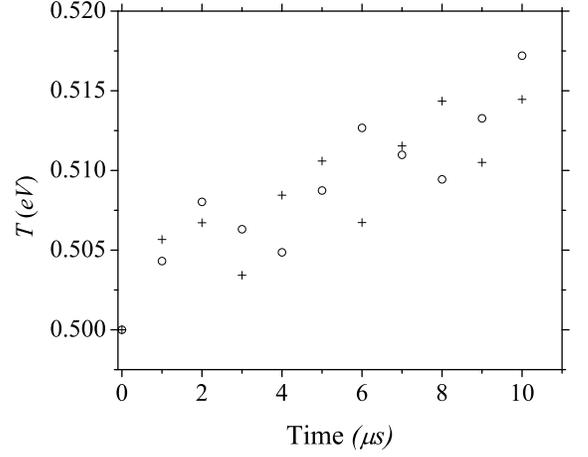}
\caption{\label{T1cm5mm}Similar trend of the plasma temperature evolution for the case W4, $\circ$, and the case W5, $+$. The numerical heating was ruled out.}
\end{figure}

It has been found in experiments that the radial transport (confinement time) is independent of pressure when the pressure is below $10^{-7}\, Torr$, and it exhibits scaling almost as ${L_p}^{-2}$ \cite{Eggleston}. The simulation results supported the idea that this "anomalous transport" is caused only by magnetostatic or electrostatic fields asymmetries as it was seen no difference on the plasma radial transport in the idealized traps in the cases W4 and W5. The anomalous loss is mainly caused by azimuthal and not longitudinal asymmetries \cite{DriscollandMalmberg}. Apparently, plasmas in microtraps experience less azimuthal magnetic field asymmetries compared to a large radius plasma in a conventional trap. On the other hand, electrostatic errors, which could arise from misalignment of the trap cylinders or sectors are more worrisome in terms of microtraps. It has been reported \cite{DriscollandMalmberg} that with improvements in trap fabrication and less misalignment ($0.1\%$), the trapped particles survived longer in experiments. This would be equivalent to $50\,nm$ precision of alignment of successive electrodes in $50\, \mu m$ radius microtrap, which is not easy to achieve by usual fabrication methods.

Other intrinsic asymmetries, such as Òpatch effectsÓ, are also present. The Òpatch effectsÓ encompass various phenomena, for instance, physically imperfect surfaces (plateaus, steps, scratches, etc.), chemical impurities, and random atomic lattice orientation, which give rise to boundary regions. These all result in a variation of the local surface work function \cite{J.F.Jia} and induce local electric fields, which can influence the charged particles and might play an important role especially when the walls get very close to the particles. With a work function variation ($\Delta\phi$) of less than $1\,mV$ for an evaporated gold surface, and estimation of the RMS potential variation along the axis of a cylindrical electrode, $RMS\,\Phi$, as \cite{Camp}
\begin{equation}\label{RMSdeltaV}
RMS\,\Phi=\frac{0.6\,\Delta\phi \,l_{c}}{R_w},
\end{equation}
it is calculated that $RMS\,\Phi<10^{-6}\,V$, when $R_w = 50\,\mu m$ and patch length, $l_c$, is comparable to the grain size of sputtered gold onto a silicon made microtrap,  $0.1\,\mu m$ \cite{Trap}.

The perpendicular drift velocity of a positron due to the patch field can be assumed as
\begin{equation}\label{patch1}
V_{\bot}=\frac{E}{B},
\end{equation}
and also the movement as
\begin{equation}\label{patch2}
\Delta x=V_{\bot} \frac{l_{c}}{V_0},
\end{equation}
where $V_0$ is the velocity by which the positron passes over the patch length (almost equal to the total velocity in a high magnetic field). The movement due to $N$ equal patches can be written then as
\begin{equation}\label{patch3}
\overline{X}^2=N{(\Delta x)}^2.
\end{equation}
Since $Nl_c=V_0t$, by substituting $N$ and $\Delta x$ in Eq. (\ref{patch3}) we obtain $t$ as
\begin{equation}\label{patch4}
t=\frac{B^2 {\overline{X}}^2 V_0}{E^2 l_c}.
\end{equation}
One calculates $t\approx4000\,s$, the time for the positron to get from the microtrap axis to the gold coated wall when $\overline{X}=R_w=50\,\mu m$, $B=7\,T$, $V_0= 1.32\times 10^{6}\,ms^{-1}$ for a $5\,eV$ positron, $l_c=0.1\,\mu m$, and $E=(1\,mV)/(50\,\mu m)=20\,Vm^{-1}$.

More realistic effect of these stray electric fields on the lifetime of a confined particle ensemble in the plasma regime could be a subject of further research while it is not expected to be a dominant factor since variations in the tube radius, which are about $\approx 1\,\mu m$ with the current fabrication process, play a bigger role than that calculated from Eqs. (\ref{RMSdeltaV}) and (\ref{patch4}).

\subsubsection{Maximum axially confined density}
We have studied so far the behavior of a plasma with  $3.75\,V$ space charge potential on axis. One would also like to know what the highest density of positrons is which is to be confined in the  $50\,\mu m$ radius microtrap with the  $10\,V$ end electrodes. In order to investigate this, the microtrap was filled up initially with a uniform plasma of higher densities corresponding to the space charge of  $37.5\,V$, $75\,V$, and $150\,V$ on axis in the cases W6, W7, and W8, respectively. For example, suppose that the end electrodes potential were much higher at the first which let us to fill the microtrap up to these initial densities. Then the barriers potential was dropped to  $10\,V$. Lowering the end electrodes potentialend electrode is a well known technique to manipulate a desirable plasma with narrow energy character \cite{Weber}.

The barriers obviously were not able to trap these high space charge plasmas and particles started to escape from the end electrodes until a confineable density was reached. Figure \ref{MAXdensity1} shows the number of trapped positrons in these cases. The inner graph is a zoomed in view in which the data are fitted well with third order exponential decay function. After a sudden drop in early stages of simulation, they were mainly the higher energy particles at the tail of the Boltzmann energy distribution escaping from the end electrodes. Evaporation of high energy particles led to creation of a narrower energy plasma. After  $0.5\,\mu s$, hollow plasmas were formed with very low density at the center and peaked at the edge since the particles mostly lost close to the axis where the space charge potential was highest. Even higher number of particles was trapped when the initial density was higher. Ultimately, the trapped potential along the plasma axis would be as high as $10\,V$ for a cold plasma, as shown in Fig. \ref{MAXdensity6}. However, a hollow plasma column is not stable to diocotron modes \cite{DavidsonBook} and plasma evolves through a turbulent-like evolution at very longer times. Collisions between the particles also affect this long time evolution.

\begin{figure}[!h]
\includegraphics[width=90mm]{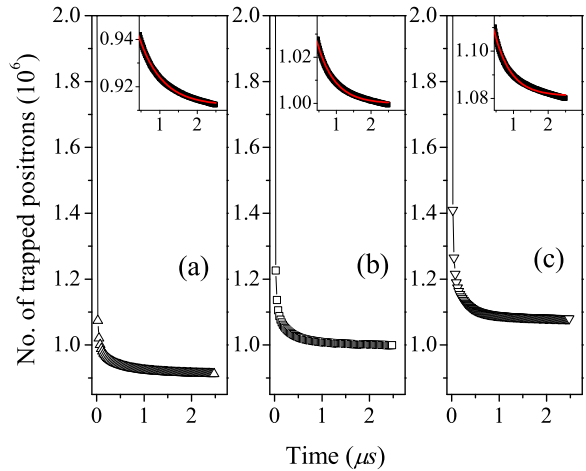}
\caption{\label{MAXdensity1}The time histories of the number of trapped positrons in (a) the case W6, (b) the case W7, and (c) the case W8. The inner graphs are zoomed in view. The data are fitted, {\color{red}$-$}, with third order exponential decay function. $36.82\%$, $20.20\%$ and $10.90\%$ of initial particles were trapped at the longer times, respectively, corresponding to densities of $5.89\times10^{11}$, $6.46\times10^{11}$  and $6.97\times10^{11}\,cm^{-3}$ in one microtrap.}
\end{figure}

\begin{figure}[!h]
\includegraphics[width=75mm]{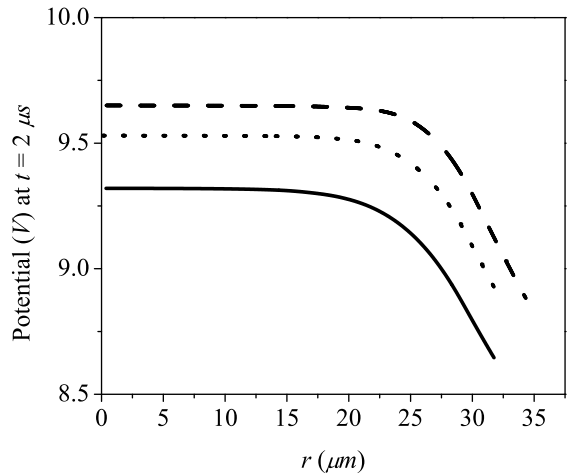}
\caption{\label{MAXdensity6}The $\phi -r$ phase at $t = 2\,\mu s$, while the initial space charge potential on axis was equal to $37.5\,V$ at the case W6,  $-$, $75\,V$ at the case W7, $...$, and $150\,V$ at the case W8, $--$.}
\end{figure}

\subsubsection{Maximum radially confined density}
For any initial state of the plasma, the axial confinement can be achieved by simply increasing the end electrodes potential. But it is the radial confinement that is worrisome. }We can easily increase the barriers potential more than one order of magnitude higher than the original $10\,V$ and still we have a quite portable trap. We simulated the case W9 and W10 with one order and two orders of magnitude higher densities from the initial analytical curve in Fig. \ref{analytical} and the end electrodes potentials of $50\,V$ and $500\,V$, respectively. Similar results of density profile and velocity phases were obtained for the case W9, and all the density was trapped. The analytical expectation from Eq. (\ref{analyticaleq}) shown in Fig. \ref{analytical} is shifted up when the space charge is increased to  $37.5\,V$, suggesting that the Brillouin limit may be surpassed at $3\,\mu m$ radius microtrap, as shown in Fig. \ref{higherend4}, while the end electrodes potential is only  $50\,V$.

\begin{figure}[!h]
\includegraphics[width=75mm]{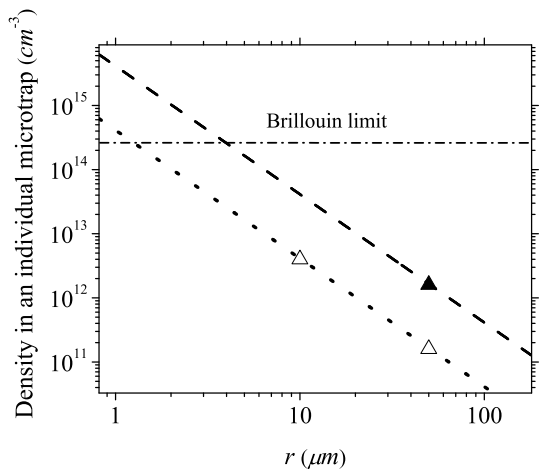}
\caption{\label{higherend4}The density of the plasma as a function of the radius of the microtrap when $R_w/R_p=\sqrt{3}$ . The analytical expectations from Eq. (\ref{analyticaleq}) are shown when the initial space charge is  $3.75\,V$ and end electrodes potential is  $10\,V$, the dot line, and when the initial space charge is  $37.5\,V$ and end electrodes potential is  $50\,V$, the dashed line. The cases W4 and W11 results, $\vartriangle$, in which the initial space charge was  $3.75\,V$ and end electrodes potential was  $10\,V$, and the case W9 result, $\blacktriangle$, in which the initial space charge was  $37.5\,V$ and end electrodes potential was  $50\,V$.}
\end{figure}

However, different results were attained for the case W10 where the initial density was even higher, $20\%$ of Brillouin limit density. The density was that high that the force balance, Eq. (\ref{forcebalance}), was not satisfied before the required amount of rotation frequency was reached to build up the inward Lorentz force. The plasma expanded to the walls which also lowered the outward electric force. About $6\%$ of initial particles hit the walls by $t=2\,\mu s$ and the exponential decay fit of trapped particles curve suggested that the $92.3\%$ of initial particles would be trapped at longer times.

\subsection{Smaller radii microtraps}

\begin{figure}[!h]
\includegraphics[width=75mm]{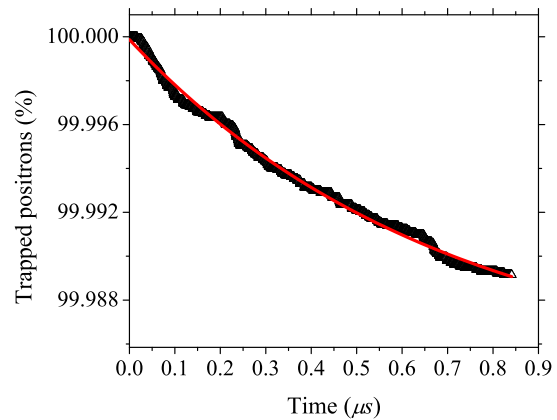}
\caption{\label{10um4}The time histories of the number of trapped positrons in the case W11. The data are fitted, {\color{red}$-$}, with a third order exponential decay function, presenting that $99.98\%$ of initial particles  were trapped at longer times.}
\end{figure}

Similar to  $50\,\mu m$ radius microtrap, the numerical heating was minimized by choosing proper parameters at the case W11 (e.g. the positron weight equal to 1). Figure \ref{10um4} illustrates the number of trapped positrons as a function of time where $99.98\%$ of initial particles  were trapped at longer times.

\section{Charged Particle Optics (CPO) simulation}
The CPO program was used to investigate the trajectories of positrons in the proposed microtrap. It uses single charge particle optics and includes space charge effects. This program uses the 'Boundary Element Method' or 'Surface Charge Method' to obtain the potential and fields at any point. CPO can incorporate space charge either by the 'space-charge cell' method or by the 'space-charge tube' method, which follow the path of select particles. The program distributes the desired charge uniformly in each cell or tube. The 'tube' method is primarily intended for simulations involving long thin beams. With the 'cell' method an accurate treatment might require that the cell dimension be too much short \cite{cpo}. For long thin beams the creation of a large number of cells is undesirable compared to a small number of tubes used in the 'tube' method. In the present simulations, the 'space-charge tube' method is used. Each ray represents a specific number of charged particles. The number of rays is limited by the length of the ray paths and the magnitude of the step time. A larger number results in more homogeneous space charge distributions. Forty-nine and sixty-four space charge tube rays were applied for the short simulations (cases S6-S8). The space charge tube diameter was set to recommended one quarter of the microtrap radius \cite{cpo}. Each iteration is defined as a complete loop in which the rays fly one time back and forth in whole length of the trap. The code uses several iterations by applying the space charge of the current iteration upon the rays for the next iteration. The motion of the charged particles was traced by investigating the trajectories of a number of rays.

For the short simulations, all rays initialized parallel to the axis of the microtrap at the center of the microtrap on a vertical middle plane. Each ray completed an iteration and ended at the same middle plane. A loop program using the $C^{++}$ language was used to setup the initial conditions of each ray in one iteration based on the final conditions of all the rays in previous iterations. If more than one ray is used, the rays are started uniformly with the same kinetic energies. This is a large simplification because the CPO program does not read in the initial conditions of all rays and each individual ray cannot be traced correctly through all iterations. At every iteration all rays were distributed within a circle. The radius of this circle and the energy of the rays were calculated from the average final radii and energies for the last iteration. As simulation progressed, some of the rays hit the microtrap wall and were lost. This process continued until the space charge decreased to a value that rays would no longer be lost. Then, the number of trapped positrons was calculated as the flight time of the remaining rays multiplied by the total current of the rays. All the short simulations based on the explained procedure are listed in Table III.

For the microtrap of $50\,\mu m$ radius, a different method was used to obtain more accurate results. The parameters used in this simulation are listed in Table III. One ray was flown along the axis of the trap from one end of the central tube to the other end, which produced a uniform cylindrical hard edge charge cloud with the radius equal to $R_w/\sqrt3$ (Space charge tube diameter, $SCTD = 2R_w/\sqrt3= 57.7\,\mu m$). This is the space charge due to the predetermined number of positrons in the trap. Individual rays were subsequently flown within this uniform, constant space charge and their trajectories were traced. The improvement of this method over that discussed above, used to obtain the short runs results presented in Fig. \ref{analytical}, is that each ray was traced correctly and without discrete iterations. In a uniform cylindrical plasma, a particle at the cloud edge experiences highest electric field. It was assumed that if one ray was flown near the cloud edge, which in our cases was radius of $28.82\,\mu m$, and did not expand in time, the whole particles would stay together without expansion. Therefore, the pre-established charge was trapped.

\begin{figure}[!h]
\includegraphics[width=75mm]{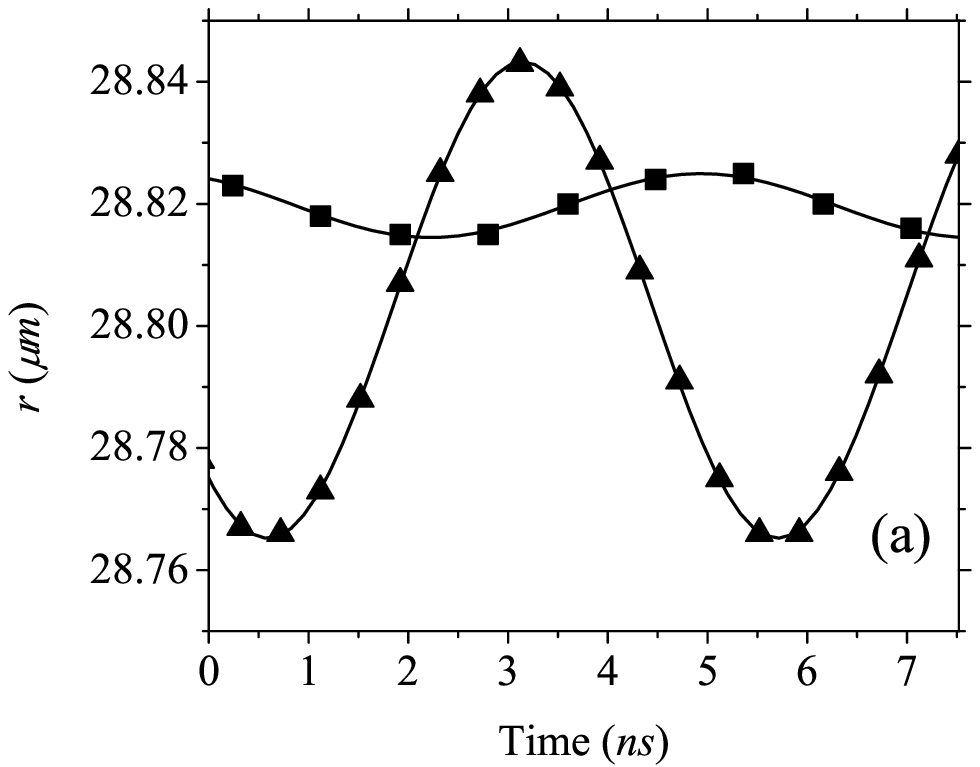}
\includegraphics[width=75mm]{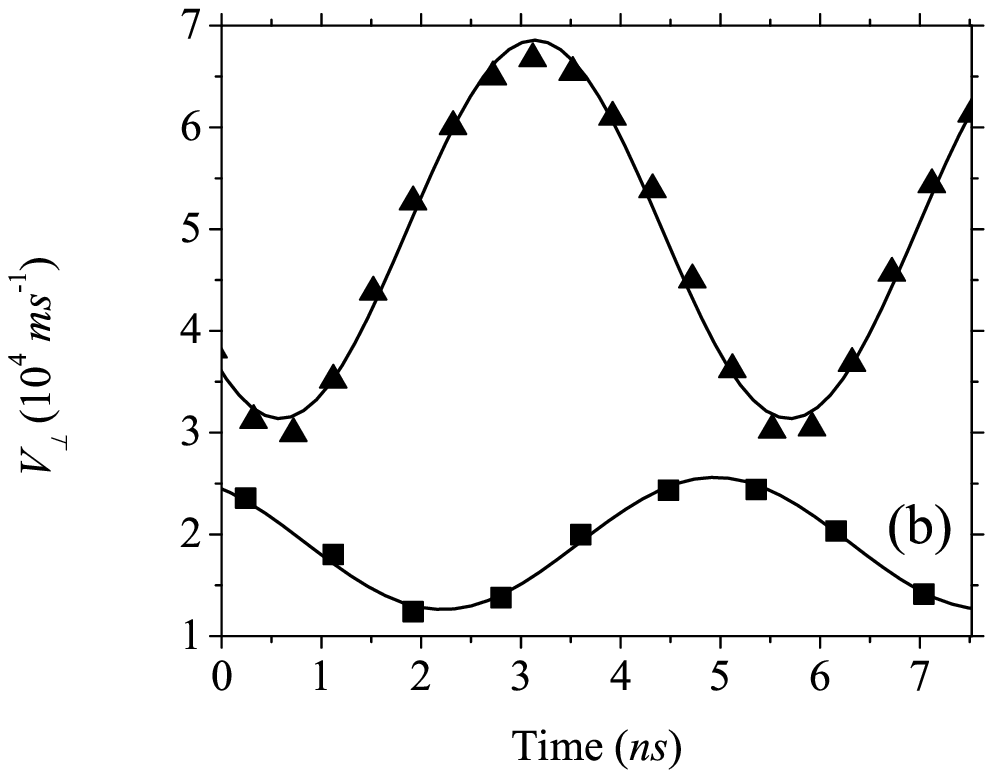}
\caption{\label{CPO1}Trajectory of a positron flying at the edge of the uniform plasma with  $4.5\,V$ of axial space charge potential with two different time steps,  $0.4\,ps$, $\blacktriangle$, and  $0.8\,ps$, $\blacksquare$. The data are fitted with equation $R(or\;V_\perp)=p_1+p_2 \sin(p_3t+p_4)$. (a) The radius data calculates  $p_1=28.80424\pm0.00007$,  $p_2=0.03901\pm0.00001$,  $p_3=982.82\pm0.86$  when step time is  $0.4\,ps$, and calculates  $p_1=28.81972\pm212.43$,  $p_2=0.00524\pm0.00016$,  $p_3=926.81\pm8.52$ when step time is  $0.8\,ps$. (b) The transverse velocity data calculates  $p_1=19113.29\pm159.58$ when step time is  $0.4\,ps$, and $p_1=49976.49\pm320.43$ when step time is  $0.8\,ps$.}
\end{figure}

In CPO, it was necessary to specify a time step that was shorter than the cyclotron period; otherwise, the trajectory integration routine did not give accurate results. The CPO routine uses the Bulirsch-Stoer method and a time step equal to the cyclotron period is insufficient \cite{cpo}. To investigate the effect of step time, a plasma was established, as described above, with radius $28.85\,\mu m$ and an axial space charge potential $4.5\,V$, in which a positron was flown at a radius $28.80\,\mu m$ with $V_z = 1.32\times10^6\,ms^{-1}$ within the central length of the microtrap. For example, when a $5\,ps$ time step was chosen, this positron lost $99\%$ of its axial velocity, which was transferred to radial velocity due to the numerical inaccuracies, within $100\,ns$. To obtain the correct helical motion, the time step must be less than $0.4\, ps$ in the case of a $7\,T$ magnetic field. Figure \ref{CPO1} shows the trajectory of a particle in presence of a uniform cloud at the cloud edge within the central length of the microtrap for time steps of $0.4$ and $0.8\, ps$. The data are fitted with equation $R(or\;V_\perp)=p_1+p_2 \sin(p_3t+p_4)$. While the mean transverse velocities ($p_1$) for the cases with time steps of $0.4\,ps$ and $0.8\,ps$ were $4.9\times10^4\,ms^{-1}$ and $1.9\times10^4\,ms{-1}$, respectively, the cyclotron radius, $r$, was obtained from
\begin{equation}\label{cyclotronradius}
r_c=\frac{mV_\perp}{q\left|\vec{B} \right|},
\end{equation}
where $V_\perp$ denotes the transverse velocity, resulting in radii of $39.6\,nm$ and $15.74\,nm$ for $0.4\, ps$ and $0.8\,ps$, respectively. The CPO simulated the radii ($p_2$) of $39.51\,nm$ and $5.50\,nm$ for these two cases consequently, maintaining $99.4\%$ and $34.9\%$ accuracy for the $0.4\,ps$ case and the $0.8\,ps$ case in the order given. Note that the radius of the gyro-center, $p_1$, in the $0.4\,ps$ case was smaller than the case with $0.8\,ps$ step time. While the initial radius of the positron was $28.80\,\mu m$, it implies that the positron in the case with larger step time experienced $5$ times higher expansion rate. Simulation resulted in a cyclotron period, $2\pi/p_3$, equal to $6.4\,ps$ and $6.8\,ps$ using step times $0.4\,ps$ and $0.8\, ps$, respectively.

For a plasma with  $0.5\,eV$ temperature, the majority of particles have kinetic energy less than few $eV$. Since it can also be assumed that, initially, the most energetic particles are more likely to expand, a particle with  $5\,eV$ kinetic energy was chosen in CPO simulations.

\begin{figure}[!h]
\includegraphics[width=75mm]{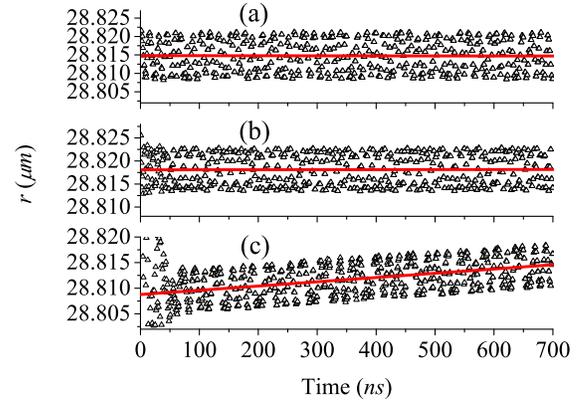}
\caption{\label{CPO2}Radius of the positron flying in presence of the uniform hard edge plasma with $28.85\,\mu m$ radius and the axial space charge potential of (a) $0.375\,V$ in the case C1, (b) $1\,V$ in the case C2, and (c) $3.75\,V$ in the case C3. The initial radius of the positron is $28.82\,\mu m$. The data are fitted with the linear function, {\color{red}$-$}, which calculates radius change rate of $-0.13\pm0.60$ for the case C1 with $\varphi_0= 0.375\,V$, $0.08\pm0.46$ for the case C2 with $\varphi_0= 1\,V$, and $6.92\pm0.3$ for the case C3 with $\varphi_0= 3.75\,V$.}
\end{figure}

Different gap sizes between grounded central tube and end electrodes were studied while in the gap region it was defined either a linearly changed voltage electrode, or overlapped electrodes of different radii and overlapping sizes, or even no electrode. The results from CPO were not consistent when the parameters of the end electrodes and gaps were slightly varied. Another problem of modeling end electrodes was that as the ray comes near them it slows down, stops and then returns on axis while the program assigns a charge   uniformly distributed along the step line, where $q=I\times \Delta t$ . So the space charge was deposited uniformly on time steps, causing high space charge accumulation near the end electrodes, which was highest at the turning point. The positrons which entered that high potential were bounced back. In order to simplify the simulation for a microtrap which has a high aspect ratio, we assume that the particle flies within an infinitely long microtrap. In the interest of study the effect of the space charge on the trajectory of one positron that is not affected by the end electrodes, a shell $C^{++}$ program was written to make the particle travel back and forth within the central region of the modeled microtrap. It stopped the particle when it reached a x-y plane located  mm away from each end electrode, recorded the data and reinitialized the particle with the same parameters but opposite axial velocity. Three cases (C1, C2 and C3) with cylindrical, uniform plasmas, of radius  $28.85\,\mu m$ and axial space charge potentials of  $0.375$, $1$  and $37.5\, V$ were simulated in the  $50\,\mu m$ radius microtrap. One particle was flown in presence of each space charge, with an initial radius of  $28.82\,\mu m$ at the plasma edge. The particle was initialized with a kinetic energy of  $5\,eV$ and no transverse velocity. Figure \ref{CPO2} shows the radius for the particle of each case as a function of time. Note that the particle expanded at an almost constant rate in each individual case. The expansion rates showed that the maximum space charge among these cases where the radius of the guiding center of the particle was almost constant was  $1\,V$ in the case C2. The particle at the edge of a space charge cloud, which experienced the highest repulsive electric field, did not move out radially if the space charge potential on the axis of the cloud was  $1\,V$ or less. At the higher density there was a clear expansion.

\section{Conclusion}
WARP results agreed well with the predictions from a simple analytic expression in terms of the axial confinement which assumes that radial confinement is satisfied below the Brillouin limit. The density was proportional to the inverse square of the trap radius. CPO results deviated dramatically and the trapped density followed more a $R_w^{-1}$  dependence.

Modeling using proper values of parameters helped to reduce the numerical heating in the WARP simulation so that the time scale in which the instability would dominate became long compared to the time required for the plasma to attain the local equilibrium. The significance of this study relies on the fact that it is given that even the initial plasma distribution with cigar shape ends was far from the equilibrium, the proposed model did result in the local equilibrium and was evolving toward the global equilibrium, while the hard edge plasma advanced to the soft edge. Final global equilibrium is to be seen in very longer time scales as radial heat transfer between the particles occurs relatively slow. One solution might be avoiding the creation of a warmer soft edge at early stages of simulation. This can be done by initializing the density profile with a relaxed edge as calculated in Fig. \ref{Math}, which remains the subject of future work. Larger computing resources and longer runs will be required to acquire a final computational equilibrium.

The length of the plasma was found to have no effect on the radial heat transfer rate and relaxation rate and so the length dependency of confinement time reported in experiments is all due to the trap and fields asymmetries. The effects of the magnetic field strength and temperature on the thermalization are also under investigation. The radial transport caused by heating due to asymmetries can be partially compensated by the resistive cooling mechanism, which is significant in micro scales \cite{Khamehchi}. However, the cyclotron radiation cooling is inhibited in this scale due to the high cut frequency of the microtrap as a waveguide.

It was demonstrated computationally by WARP code that a uniform, soft edge plasma with density of  $1.6\times10^{11}\,cm^{-3}$ can be trapped in one microtrap with the radius of  $50\,\mu m$ and confining potentials of  $10\,V$; hence $2.35\times10^{13}$  positrons can be trapped in an array of $187'500$ microtraps filling the size of a soda can ($5\,cm$ diameter and  $10\,cm$ length), assuming that the filling factor is $75\%$. This density is comparable to the highest reported density in a conventional Penning-Malmberg trap ($\approx 10^{11}\,cm^{-3}$) which uses order of kV electrostatic potentials and often some means such as rotating walls to confine the plasma radially because of the high existing space charge and outward electric forces. Ten times higher density was trapped when barrier potentials was increased to  $50\,V$, suggesting that the Brillouin limit may be surpassed at  $3\,\mu m$ radius microtrap.

CPO applies to the single particle regime and so does not include collisions between the particles. Furthermore, the results from CPO were not consistent when the parameters of the end electrodes and gaps were slightly varied. High space charge accumulation near the end electrodes was seen. Therefore, the results from this tool should be considered with caution. If the particle is low enough in energy so that it can be confined axially (e.g. a  $5\,eV$ kinetic energy particle in a plasma of  $3.75\,V$ space charge potential and a microtrap with  $10\,V$ end electrodes) one can follow the trajectory of the particle using CPO program while the end electrode issues are avoided. For the case of  $50\,\mu m$ radius traps, the corresponding density was less than a third of the density achieved with WARP. Considering the limitations of the CPO, accuracy of the results especially near the end electrodes region, and also capabilities of this program to simulate in a plasma regime, the CPO is not suitable for modeling the plasmas in PM traps or similar systems.

Simulations will be extended to smaller radius traps where the simple analytic prediction crosses the Brillouin limit. We will also try to figure out what trap radius and what aspect ratio is ideal for storing large number of particles in practice. Experimental efforts to test the long aspect ratio microtrap array are under way. Experimental and modeling results will be compared. Computational studies might also be required for the beam transport and injection into the trap.

Loses arise on experiments by patch effects, annihilation with gas molecules, and by trap imperfections such as nonalignment of microtraps, asymmetries, and non uniform magnetic field. Simulations will help to investigate these effects and find out the amount of deviations from perfectness tolerable in our design. The fabrication of microtrap arrays of $50\,\mu m$ radius and $100\,mm$ length is under way. It is achieved by deep etching $200$ silicon dies of $500\,\mu m$ thickness and $38\,mm$ diameter (each die contains $20'419$ numbers of $50\,\mu m$ holes) which are then aligned and stacked over one another to create thousands of long tubes \cite{Trap}. Coating inside the tubes with gold helps us to reduce the patch effects. With the current fabrication process, variations in the tube radius are about $2\,\mu m$, a misalignment of $4\,\mu m$ was achieved, and the scalloping size of the walls due to the Bosch process was measured about $400\,nm$. We might need to improve the etching to get more uniform holes in future.

As we go to smaller and smaller radius microtraps, the associated cloud is not indeed a plasma anymore. As a consequence the space charge potential becomes negligible in the thermal equilibrium density distribution. One can consider a nanotrap (as small as a cyclotron radius of positron) containing only one positron which avoids all plasma complications and pushes the density over the Brillouin limit, and permits confinement times limited only by vacuum conditions.

\begin{acknowledgments}
The authors are grateful to colleagues at WSU, Paola Folegati and Jia Xu for their contribution to the early stages of the project, and Randall Svancara for his helps to the simulations on HPC. We are also thankful to Dr. David Grote at LLNL for assistance and useful discussions regarding the WARP simulations, and also Dr. Frank Read for his helps toward CPO simulations. We would also like to thank program managers Dr. William Beck and Dr. Parvez Uppal of the Army Research Laboratory who provide funding under contract $W9113M-09-C-0075$, Positron Storage for Space and Missile Defense Applications, and program manager Dr. Scott Coombe of the Office of Naval Research who provide funding under award $\#N00014-10-1-0543$, Micro- and Nano-Traps to Store Large Numbers of Positron Particles at Very Large Densities.
\end{acknowledgments}

\bibliographystyle{unsrt}
\bibliography{myreference}

\end{document}